\documentclass[11pt,a4wide]{article}
\pdfoutput=1
\usepackage{jheppub}

\usepackage[english]{babel}
\usepackage[babel]{csquotes}
\usepackage{amsfonts}
\usepackage{array, tabularx}
\usepackage{mathtools}
\usepackage{amsthm}
\usepackage{url}
\usepackage{multicol, multirow}
\usepackage{emptypage}
\usepackage{verbatim}
\usepackage{dsfont}
\usepackage{hyperref}
\usepackage{float}
\usepackage{amsmath}
\usepackage{amssymb,tikz}
\usepackage{amsfonts}

\newcommand{\ahat}{\hat{a}}

\newcommand{\R}{\mathbb{R}}

\renewcommand{\hbar}{\bar{h}}
\newcommand{\hb}{\bar{h}}

\newcommand{\RR}{R}
\newcommand{\zbar}{\bar{z}}
\newcommand{\zb}{\bar{z}}
\newcommand{\rd}{{\mathrm d}}

\newcommand{\cO}{\mathcal{O}}
\newcommand{\cG}{\mathcal{G}}
\newcommand{\bT}{\mathbf{T}}

\newcommand{\degsum}[1]{\llangle #1 \rrangle}

\theoremstyle{definition}

\theoremstyle{remark}

\theoremstyle{definition}

\allowdisplaybreaks[2]

\usepackage[font=small,format=hang,labelfont={sf,bf}]{caption}

\makeatletter
\DeclareFontFamily{OMX}{MnSymbolE}{}
\DeclareSymbolFont{MnLargeSymbols}{OMX}{MnSymbolE}{m}{n}
\SetSymbolFont{MnLargeSymbols}{bold}{OMX}{MnSymbolE}{b}{n}
\DeclareFontShape{OMX}{MnSymbolE}{m}{n}{
    <-6>  MnSymbolE5
   <6-7>  MnSymbolE6
   <7-8>  MnSymbolE7
   <8-9>  MnSymbolE8
   <9-10> MnSymbolE9
  <10-12> MnSymbolE10
  <12->   MnSymbolE12
}{}
\DeclareFontShape{OMX}{MnSymbolE}{b}{n}{
    <-6>  MnSymbolE-Bold5
   <6-7>  MnSymbolE-Bold6
   <7-8>  MnSymbolE-Bold7
   <8-9>  MnSymbolE-Bold8
   <9-10> MnSymbolE-Bold9
  <10-12> MnSymbolE-Bold10
  <12->   MnSymbolE-Bold12
}{}

\let\llangle\@undefined
\let\rrangle\@undefined
\DeclareMathDelimiter{\llangle}{\mathopen}%
                     {MnLargeSymbols}{'164}{MnLargeSymbols}{'164}
\DeclareMathDelimiter{\rrangle}{\mathclose}%
                     {MnLargeSymbols}{'171}{MnLargeSymbols}{'171}
\makeatother

\title{Critical \( \mathrm O(N)\) model to order \(\epsilon^4\) from analytic bootstrap}

\author{Johan Henriksson \&}
\author{Mark van Loon}
\affiliation{Mathematical Institute, University of Oxford, Andrew Wiles Building, Radcliffe Observatory Quarter, Woodstock Road, Oxford, OX2 6GG, UK}

\emailAdd{johan.henriksson@maths.ox.ac.uk}
\emailAdd{mark.vanloon@maths.ox.ac.uk}

\abstract{
We compute, using the method of large spin perturbation theory, the anomalous dimensions and OPE coefficients of all leading twist operators in the critical \( \mathrm{O}(N)\) model, to fourth order in the \( \epsilon\)-expansion.
This is done fully within a bootstrap framework, and generalizes a recent result for the CFT-data of the Wilson--Fisher model.
The anomalous dimensions we obtain for the \(\mathrm{O}(N)\) singlet operators agree with the literature values, obtained by diagrammatic techniques, while the anomalous dimensions for operators in other representations, as well as all OPE coefficients, are new.
From the results for the OPE coefficients, we derive the \(\epsilon^4\) corrections to the central charges \( C_T \) and \( C_J \), which are found to be compatible with the known large \(N\) expansions.
Predictions for the central charge in the strongly coupled 3d model, including the 3d Ising model, are made for various values of \(N\), which compare favourably with numerical results and previous predictions.
}

\begin{document}
\maketitle


\section{Introduction}
\label{sec:intro}
A conformal field theory (CFT) can be characterized by its CFT-data, consisting of a spectrum of primary operators, described by their scaling dimensions and spins \( \{ (\Delta,\ell) \} \), and the algebra which they satisfy, known as the Operator Product Expansion (OPE).
The conformal bootstrap \cite{Rattazzi:2008pe} imposes the associativity of this algebra on specific correlators to constrain the CFT-data of exchanged intermediate operators. 

Whereas numeric bootstrap methods have mostly focused on finding CFT-data for operators with low spin, analytic methods have been developed based on properties of operators with large spin \cite{Alday:2007mf,Fitzpatrick:2012yx,Komargodski:2012ek,Alday:2015eya}. 
These developments led to the method of large spin perturbation theory \cite{Alday:2016njk}, a perturbation theory around the infinite spin point at which the theory is essentially free. 
In this method, CFT-data for spinning operators can be computed from the double discontinuity of the correlator, using either twist conformal blocks \cite{Alday:2016njk} or the recent Froissart--Gribov inversion formula \cite{Caron-Huot:2017vep,Alday:2017vkk}. 
The latter guarantees that, assuming the correct Regge behaviour, the CFT-data for \(\ell>1\) is an analytic function in the spin \cite{Caron-Huot:2017vep,Simmons-Duffin:2017nub}.
Crossing symmetry allows for the double discontinuity to be computed from the contribution of specific operators in the crossed channel.
Using this, large spin perturbation theory has been successfully applied to find CFT-data for a number of models \cite{Alday:2016jfr,Alday:2017xua,Dey:2017fab,Henriksson:2017eej,vanLoon:2017xlq,Alday:2017vkk}.
In a recent paper \cite{Alday:2017zzv}, these methods were applied to find the CFT-data of weakly broken higher spin currents in the Wilson--Fisher model in \( d = 4-\epsilon \) dimensions, to fourth order in the \( \epsilon \)-expansion, confirming some previously obtained results for anomalous dimensions, and producing new results for OPE coefficients and the central charge.
This was possible because to third order the double discontinuity receives contributions from just the identity and one other operator, while the fourth order contributions arise only from two well-behaved towers of operators of approximate twists \(\tau = 2,4\).

The object of study in this paper is the critical \(\mathrm O(N)\) model, a conformal scalar field theory with fields \(\varphi^i\) transforming in the fundamental representation of a global \(\mathrm O(N)\) symmetry. In the framework of perturbative quantum field theory, the theory has been studied by perturbing \(N\) free fields with a quartic interaction \(\lambda(\varphi^i\varphi^i)^2\). In \(d=4-\epsilon\) dimensions, this theory has an IR fixed-point with \(\lambda\sim\epsilon\). At this point the theory becomes conformal and is referred to as the \emph{critical \(\mathrm O(N)\) model}. 
Our study of the theory, however, will employ the bootstrap idea and make no reference at all to the Lagrangian description or Feynman diagrams. 
We will assume conformal and global \(\mathrm O(N)\) symmetry, and make some very general assumptions about the spectrum of operators (essentially that the spectrum is a perturbation from generalized free fields). Using crossing symmetry and general results for CFTs, we will then be able to show that the results that follow from these assumptions agree with the critical \(\mathrm O(N)\) model. 
Furthermore, our approach provides an alternative to computing CFT-data using diagrammatic methods, and produces results of higher orders in the \(\epsilon\)-expansion than previously computed. 
As an application, we will use the \(\epsilon\)-expansion results to approximate some strongly interacting critical \(\mathrm O(N)\) models in three dimensions, which describe critical systems in different universality classes, such as the XY class (\(N=2\)) for the superfluid transition of \(\mathrm{^4He}\) and the Heisenberg class (\(N=3\)) for the critical behaviour of isotropic magnets\footnote{For an extensive review of the physical applications, see \cite{Pelissetto:2000ek}.}.

In \cite{Alday:2017zzv} the critical \(\mathrm O(N)\) model was studied at order \(\epsilon^3\), and in this paper we will extend the computation to compute results at fourth order. 
Since the crossing symmetry relation mixes the contributions from different \(\mathrm{O}(N)\) representations, determining the exact form of the contribution from quadrilinear operators is a non-trivial problem, not present in the \(N=1\) case. 
In particular, the existence of two bilinear singlet operators means that there will now be two different expansion parameters entering into the quadrilinear contribution.
Having resolved this problem and computed the quadrilinear contribution, we fully determine, for all spins \( \ell > 1 \), the CFT-data of the weakly broken higher spin currents in the \( \mathrm{O}(N) \) model, in the singlet, traceless symmetric, and antisymmetric \(\mathrm{O}(N)\) representations.
Subsequently, assuming that the results for the CFT-data can be analytically continued to \(\ell=0\), we constrain the parameters in our results and find that they agree with the critical \(\mathrm O(N)\) model. 
The quartic order results that we derive are new, except the anomalous dimensions in the \(\mathrm O(N)\) singlet, where we reproduce the recent diagrammatic calculation \cite{Derkachov:1997pf,Manashov:2017xtt}.
Furthermore all our results agree known results computed in the large \(N\) limit \cite{Lang:1993ge,Petkou:1995vu,Derkachov:1997ch}.
Finally we combine our results for the \(\mathrm O(N)\) model in \(d=4-\epsilon\) dimensions with the known results for the non-linear sigma model in \(d=2+\epsilon\) dimensions \cite{Diab:2016spb} to provide Pad{\' e} approximations to the central charge in the strongly interacting 3d models. 
These provide an improvement on previous predictions and show excellent agreement with numerical results.

Section \ref{sec:generalities} gives a brief overview of the general method of large spin perturbation theory in the presence of a global \(\mathrm{O}(N)\) symmetry. 
Section \ref{sec:inversion} applies this method to find the CFT-data in the \(\mathrm{O}(N)\) model, focusing on the contribution from quadrilinear operators of approximate twist 4.
In section \ref{sec:lowLyingSpin} we constrain the various parameters in the theory using matching conditions for leading twist operators of low spins, and we pay particular attention to the subtle analytic continuation of the CFT-data to spin \(\ell=0\).
In section \ref{sec:CentralCharges} we derive new predictions for the central charges \( C_T \) and \( C_J \) associated to the stress-energy tensor and global symmetry current respectively, and we use Pad{\' e} approximants to provide predictions for these charges in 3d.

Appendix \ref{app:subleadBlocks} provides some useful background on the small-\(z\) behaviour of conformal blocks that is used in our computation.
Appendices \ref{app:I2andI4} and \ref{app:CFTdata} present the full results of our computation, which can also be found in the Mathematica file included with this paper.


\section{Large spin perturbation theory with global symmetry}
\label{sec:generalities}


We study the correlator of four fundamental fields in general dimension \( d \) in the presence of a global \(\mathrm{O}(N)\) symmetry
\begin{equation}
\langle \varphi_i(x_1) \varphi_j(x_2) \varphi_k(x_3) \varphi_l(x_4) \rangle = \frac{\cG_{ijkl}(z,\zbar)}{x_{12}^{2\Delta_\varphi}x_{34}^{2\Delta_\varphi}},
\end{equation}
where the standard conformal cross-ratios are
\begin{equation}
u = z \zbar = \frac{x_{12}^2 x_{34}^2}{x_{13}^2 x_{34}^2},\qquad 
v =(1-z)(1-\zbar) =  \frac{x_{14}^2 x_{23}^2}{x_{13}^2 x_{34}^2},
\end{equation}
and where \( \cG_{ijkl}(z,\zbar) \) satisfies crossing symmetry:
\begin{equation} \label{eq:genCrossing}
\left(\frac{1-z}{z} \right)^{\Delta_\varphi}\cG_{ijkl}(z,\zbar)  = \left(\frac{\zbar}{1-\zbar} \right)^{\Delta_\varphi} \cG_{kjil}(1-\zbar,1-z).
\end{equation}
The intermediate operators transform in the singlet (\(S\)), traceless symmetric (\(T\)) and antisymmetric (\(A\)) representations of the global \( \mathrm{O}(N) \) symmetry, and as such the correlator decomposes
\begin{equation}
\cG_{ijkl}(z,\zbar) = \cG_S(z,\zbar) \bT_{ijkl}^S + \cG_T(z,\zbar) \bT_{ijkl}^T + \cG_A(z,\zbar) \bT_{ijkl}^A\, ,
\end{equation}
where we defined the following basis of tensor structures
\begin{equation}
\bT^S_{ijkl} = \delta_{ij}\delta_{kl},\qquad 
\bT^T_{ijkl} = \frac{\delta_{ik}\delta_{jl} + \delta_{il}\delta_{jk}}{2} - \frac{1}{N} \delta_{ij}\delta_{kl}, \qquad
\bT^A_{ijkl} = \frac{\delta_{ik}\delta_{jl} - \delta_{il}\delta_{jk}}{2}.
\end{equation}
The four-point functions can be decomposed into conformal blocks as follows
\begin{equation}
\cG_\RR(z,\zbar) = \sum_{\cO_{\RR,\Delta,\ell}} a_{\RR,\Delta,\ell} (z\zbar)^{\tau/2} G^{(d)}_{\Delta,l}(z,\zbar),
\end{equation}
where the sum is over all intermediate operators \( \cO_{\RR,\Delta,\ell} \) in the respective \(\mathrm O(N)\) representation, parametrised by scaling dimension \(\Delta\) and spin \(\ell\).
 In this equation, \(a_{\RR,\Delta,\ell} \) is the relevant (squared) OPE coefficient, \(\tau=\Delta-\ell\) is the twist, and \( G^{(d)}_{\Delta,\ell}(z,\zbar) \) is the \(d\)-dimensional conformal block.

The crossing relation \eqref{eq:genCrossing} decomposes as
\begin{align} \label{eq:ONcrossing}
\cG_S(u,v) &= \left(\frac{u}{v}\right)^{\Delta_\varphi}\left( \frac{1}{N}\cG_S(v,u) + \frac{(N+2)(N-1)}{2N^2} \cG_T(v,u) +\frac{1-N}{2N} \cG_A(v,u)\right), \nonumber\\
\cG_T(u,v) &=\left(\frac{u}{v}\right)^{\Delta_\varphi}\left( \cG_S(v,u)+ \frac{N-2}{2N} \cG_T(v,u) +\frac{1}{2} \cG_A(v,u)\right), \\
\cG_A(u,v) &=\left(\frac{u}{v}\right)^{\Delta_\varphi}\left( -\cG_S(v,u) + \frac{N+2}{2N} \cG_T(v,u) +\frac{1}{2} \cG_A(v,u)\right). \nonumber
\end{align}

We shall establish an \(\epsilon\)-expansion around the generalized free theory in \( d = 4- \epsilon \) dimensions.
The correlator in the generalized free theory is simply
\begin{equation}
\cG_{ijkl}^{(0)}(z,\zbar) = \delta_{ij}\delta_{kl}+ u^{\Delta_\varphi} \delta_{ik}\delta_{jl} + \left(\frac{u}{v}\right)^{\Delta_\varphi} \delta_{il}\delta_{jk},
\end{equation}
which implies that
\begin{align}
\cG^{(0)}_S(z,\zbar) &= 1+\frac{1}{N}u^{\Delta_\varphi}\left(1+\frac{1}{v^{\Delta_\varphi}}\right),\nonumber\\
\cG^{(0)}_T(z,\zbar) &= u^{\Delta_\varphi}\left(1+\frac{1}{v^{\Delta_\varphi}}\right),\\
\cG^{(0)}_A(z,\zbar) &= u^{\Delta_\varphi}\left(1-\frac{1}{v^{\Delta_\varphi}}\right).\nonumber
\end{align}
The intermediate operators have dimensions \( \Delta_{n,\ell} = 2\Delta_\varphi + 2n + \ell \) for \(n=0,1,\ldots\) .
The OPE coefficients \( a_{T,\Delta,\ell} \) of operators in the traceless symmetric representation match the OPE coefficients in the \( \langle \varphi\varphi\varphi\varphi \rangle \) correlator in the theory of a single generalized free field; for this reason we shall refer to them as the generalized free field OPE coefficients \( a^{GFF}_{\Delta,\ell} \).
They are (see e.g. \cite{Gliozzi:2017hni, Fitzpatrick:2011dm}\footnote{Different normalizations of conformal blocks, depending on whether one includes factors of the form \(2^{-\ell}\) and \(\frac{(d/2-1)_\ell}{(d-2)_\ell}\), are used in \cite{Gliozzi:2017hni} and \cite{Fitzpatrick:2011dm}, both of which differ from the normalization of the blocks in this note. })
\begin{equation} \label{eq:GFFOPECoefficients}
a^{GFF}_{\Delta_{n,\ell},\ell} = \frac{2(\Delta_\varphi-\frac{d}{2}+1)_n^2 (\Delta_\varphi)_{\ell+n}^2}{\ell!n! (\frac{d}{2}+\ell)_n (2 \Delta_\varphi+n-d+1)_n (2 \Delta_\varphi+ 2n+\ell  -1)_\ell (2 \Delta_\varphi+n+\ell-\frac{d}{2})_n},
\end{equation}
for even \( \ell \), with the odd spin OPE coefficients vanishing.
Here \( (a)_b\equiv \Gamma(a+b)/\Gamma(a) \) is the Pochhammer symbol. 
The OPE coefficients of operators in the antisymmetric representation are of the exact same form, except that the formula is valid only for odd \( \ell\) and that the coefficients vanish for even \( \ell \).

Perturbing to the interacting \( \mathrm{O}(N) \) model breaks higher spin symmetry, and the intermediate operators acquire corrections.
As in \cite{Alday:2017zzv}, we define anomalous dimensions and OPE coefficient corrections  with respect to their generalized free field values.
Furthermore it is known that at first order in \( \epsilon\), crossing symmetry constrains the CFT-data to consist only of a finite support solution at spin \( \ell = 0 \) \cite{Alday:2016jfr}.
Therefore, amongst the intermediate operators of leading twist, only the scalar operators acquire first order anomalous dimensions. 
Such operators exist in the \(\mathrm O(N)\) singlet and traceless symmetric representations and
we shall denote the corrections to their anomalous dimensions by \(g_S\) and \( g_T\) respectively.
Thus we define
\begin{equation}\label{eq:definitionOfgSgT}
\Delta_{\varphi_S^2} = 2\Delta_\varphi + g_S, \qquad 
\Delta_{\varphi_T^2} = 2\Delta_\varphi + g_T,
\end{equation}
and use \(g_S\) and \(g_T\) as perturbative expansion parameters, which is possible since \( g_S, g_T \sim \epsilon\).

To find the CFT-data of the leading twist operators, we take the conformal block expansion of the correlators \( \cG_{\RR}(z,\zbar)\), expand as series in \( \epsilon \) up to order \( \epsilon^4\), and take the limit of small \(z\), to reduce the crossing equations \eqref{eq:ONcrossing} to
\begin{equation} \label{eq:GenInvProblem}
\sum_\ell z^{\Delta_\varphi} \left( U_{\RR,\hbar}^{(0)} + \frac{1}{2} \log z \,U^{(1)}_{\RR,\hbar}  + \frac{1}{8}\log^2 z\, U^{(2)}_{\RR,\hbar} \right) f_{\Delta,\ell}(\zbar) = Q_{\RR}(z,\zbar),
\end{equation}
where each \( Q_{\RR}(z,\zbar) \) is the small-\(z\) limit of the right-hand side of the crossing equation \eqref{eq:ONcrossing} for representation \( \RR\), up to order \( \epsilon^4 \).
We also defined \( \hbar \equiv \ell + \Delta_\varphi \), and defined the functions \( f_{\Delta,\ell}(\zbar) \) to be the \(\mathrm{SL}(2,\R) \) conformal blocks, normalized as follows:
\begin{equation}
f_{\Delta,\ell}(\zbar) = r_{\frac{\Delta+\ell}{2}} k_{\frac{\Delta+\ell}{2}}(\zbar), \qquad r_h = \frac{\Gamma(h)^2}{\Gamma(2h)},
\end{equation}
with \( k_\beta(\zbar) = \zbar^\beta \,_2F_1(\beta,\beta;2\beta;\zbar)\).
Furthermore the \( U^{(m)}_{\RR, \hbar} \) are related to the CFT-data as follows \cite{Alday:2017vkk, Alday:2017zzv}:
\begin{equation} \label{eq:DefnOfU}
\ahat_{\RR}(\hbar) \left( \gamma_{\RR,\ell} \right)^p = U^{(p)}_{\RR,\hbar} + \frac{1}{2} \partial_{\hbar} U^{(p+1)}_{\RR,\hbar}+ \frac{1}{8} \partial^2_{\hbar} U^{(p+2)}_{\RR,\hbar} + \dots ,
\end{equation}
where the \( \ahat_{\RR}(\hbar) \) correspond to the OPE coefficients as \( a_{\RR,\Delta,\ell} = r_{\frac{\Delta+\ell}2} \ahat_{\RR}(\hbar)\).
The problem of finding the \(U^{(p)}_{\RR,\hbar} \) from \eqref{eq:GenInvProblem}, given a known \( Q_{\RR}(z,\zbar) \), was discussed in \cite{Alday:2017zzv} and we do not repeat the discussion here.
It has the following solution: if
\begin{equation}
\sum_{\Delta=2\Delta_\varphi+\ell} \ahat(\hbar) f_{\Delta,\ell}(\zbar) = G(\zbar), 
\end{equation}
then 
\begin{equation}
\ahat(\hbar) = \frac{2 \hbar-1}{\pi^2} \int_0^1 \rd t\int_0^1 \rd\zbar ~\frac{\zbar^{\hbar-2}(t(1-t))^{\hbar-1}}{(1-t \zbar)^{\hbar}}  {\rm dDisc} \left[G(\zbar)\right],
\end{equation}
where the double discontinuity of a correlator is defined as the difference between the Euclidean correlator and its two analytic continuations around \( \zbar=1 \):
\begin{equation}
{\rm dDisc}\,[G(\zbar)] \equiv G(\zbar) -\frac{1}{2} G^\circlearrowleft(\zbar)-\frac{1}{2} G^\circlearrowright(\zbar).
\end{equation}
The appendix of \cite{Alday:2017zzv} gives the result of applying the formula to various functions \(G(\zbar)\) encountered in this paper.


\section{Inversion of the \(\mathrm{O}(N)\) model}
\label{sec:inversion}


Let us now discuss the precise inversion problem we solve.
Take for instance the first crossing relation in \eqref{eq:ONcrossing},
\begin{equation}
\cG_S(u,v) = \left(\frac{u}{v}\right)^{\Delta_\varphi}\left( \frac{1}{N}\cG_S(v,u) + \frac{(N+2)(N-1)}{2N^2} \cG_T(v,u) +\frac{1-N}{2N} \cG_A(v,u)\right),
\end{equation}
and consider the terms on the right-hand-side that create a double discontinuity.
Taking the small-\(z\) limit, we arrive at the following inversion problem for singlet operators of leading twist:
\begin{equation} \label{eq:ONSingInv}
\sum_{\Delta_{S,\ell} = 2\Delta_\varphi+\ell} \hat a_\ell z^{\tau_{S,\ell}/2} f_{\Delta_{S,\ell},\ell}(\zbar) 
=  \left. z^{\Delta_\varphi} \left(\frac{1}{N}D_S + \frac{(N+2)(N-1)}{2N^2} D_T +\frac{1-N}{2N} D_A\right)\right|_{\mathrm{small}\,z}
\end{equation} 
where we have defined
\begin{equation}
D_\RR =  \mathrm{dDisc}\left[\left(\frac{\zbar}{1-\zbar}\right)^{\Delta_\varphi} \cG_\RR(1-\zbar,1-z) \right].
\end{equation}
To order \( \cO(\epsilon^4) \), the double discontinuities take the form
\begin{align}
D_S &= \mathrm{dDisc}\Big[I_{\mathds{1}} + I_{\varphi^2_S} + I^S_2 + I^S_4 \,\Big],\nonumber\\
D_T &= \mathrm{dDisc}\Big[\qquad\, I_{\varphi^2_T} + I^T_2 + I^T_4 \Big],\label{eq:DoubleDiscontinutities}\\
D_A &= \mathrm{dDisc}\Big[\qquad\qquad\quad I^A_2 + I^A_4 \Big].\nonumber
\end{align}
Here \( I_{\mathds{1}} \) of order \( \cO(\epsilon^0) \) and \( I_{\varphi^2_S}\), \(I_{\varphi^2_T}\) of order \( \cO(\epsilon^2)\) correspond to the exchange of the identity operator and the bilinear scalars in the crossed channel, whereas
\( I_2^R\) and \(I_4^R\) are of order \( \cO(\epsilon^4)\) and come from infinite towers of operators of approximate twists \(2\) and \(4\) respectively. 
At this order in \(\epsilon\), there are no contributions from operators of higher twist.
Let us discuss in more detail the precise forms of the discontinuities in \eqref{eq:DoubleDiscontinutities}.

\subsection{Contributions to order \( \epsilon^3\)} \label{sec:uptocubicorder}
At zeroth order in \( \epsilon \), there is only the double discontinuity \( I_{\mathds{1}} \) from the identity operator in the cross-channel:
\begin{equation}
I_{\mathds{1}} = \left(\frac{\zbar}{1-\zbar}\right)^{\Delta_\varphi}.
\end{equation}

At second order in \(\epsilon\), the only new contributions to the double discontinuity come from scalar operators in the singlet and traceless symmetric representations, whose anomalous dimensions \(g_S\) and \(g_T\) \eqref{eq:definitionOfgSgT} are of order \(\epsilon\).
For example,
\begin{align}
I_{\varphi^2_S}&=
\left(\frac{\zbar}{1-\zbar}\right)^{\Delta_\varphi} a_{S,0}\, (1-\zbar)^{\frac{1}{2}\Delta_{S,0}} G^{(d)}_{\Delta_{S,0},0}(1-\zbar,1-z) + \cO(\epsilon^4) \nonumber\\\label{eq:Ivarphi2S}
&= \zbar^{\Delta_\varphi} a_{S,0}\, (1-\zbar)^{g_S/2} G^{(d)}_{\Delta_{S,0},0}(1-\zbar,1-z) + \cO(\epsilon^4),
\end{align}
where \(\Delta_{S,0} =\Delta_{\varphi^2_S}
\)
The terms with a non-zero double discontinuity arise in two ways: firstly from expanding the term \( (1-\zbar)^{g_S/2}\) in powers of \(g_S\), and secondly from Taylor expanding the conformal block 
\begin{equation}\label{eq:scalarBlockS}
G^{(d)}_{\Delta_{S,0},0}(1-\zbar,1-z)=G^{(4-\epsilon)}_{2\Delta_\varphi+g_S,0}(1-\zbar,1-z).
\end{equation}
Notice that the conformal block \(G_{\Delta,0}^{(d)}(1-\zbar,1-z)\) itself is regular as \(\zbar\rightarrow1\), and therefore the double-discontinuity of \(I_{\varphi^2_S}\) will start at order \(g_S^2\). This is a general feature that explains the orders at which the double discontinuities \eqref{eq:DoubleDiscontinutities} appear; the contributions in the crossed channel from operators with dimensions \(2\Delta_{\varphi}+n+\gamma\) for integer \(n\) have a double discontinuity that starts at order \(\gamma^2\).
To evaluate the expansion \eqref{eq:scalarBlockS} of the conformal blocks, we use the precise form of the \(d\)-dimensional conformal block for the scalar exchange between two identical scalar external operators \cite{Dolan:2000ut}:
\begin{equation}
G_{\Delta,0}^{(d)}(1-\zbar,1-z) = \sum_{m,n=0}^\infty \frac{\left(\Delta/2\right)^2_m\left(\Delta/2\right)^2_{m+n}}{m! n! \left( \Delta+1-d/2\right)_m \left(\Delta\right)_{2m+n}} (1-z)^m(1-\zbar)^m(1-z \zbar)^n,
\end{equation}
By first performing the sum over \(n\) and then expansing for small \(z\), the contributions \(I_{\varphi^2_S}\) and \( I_{\varphi^2_T}\) can be readily computed. Notice that in the expression \eqref{eq:Ivarphi2S}, and correspondingly for \(I_{\varphi_T^2}\), we need to allow for the OPE coefficients to acquire corrections and  therefore we write
\begin{equation}
a_{S,0}=\frac{2}{N}\left(1+\alpha_S\right),\qquad 
a_{T,0}=2\left(1+\alpha_T\right).
\end{equation}
The corrections \(\alpha_S\) and \(\alpha_T\) start at order \(\epsilon\) and in section~\ref{sec:lowLyingSpin} we will be able to compute them to order \(\epsilon^2\), which is the maximal order they appear with in our \(\epsilon^4\) results.

\subsection{Contributions to order \( \epsilon^4\)} \label{sec:fourthorder}
At fourth order in \( \epsilon \), we get two additional contributions: \( I_2^\RR\) and \(I_4^\RR\), coming from bilinear and quadrilinear operators, respectively.

\subsubsection{Leading twist operators}
Since the anomalous dimensions of the leading twist operators start at order \( \epsilon^2 \), the sums of their squares give contributions \( I_2^\RR\) to the double discontinuity at order \( \epsilon^4 \). 
Therefore, at this order, we may evaluate the conformal blocks in four dimensions, leading to, for example, the singlet contribution
\begin{equation}
I_2^S = \frac{1}{8} \log^2(1-\zb) \sum_{\ell=0,2,\dots} a_{S,\ell} \left(\gamma_{S,\ell}\right)^2 G^{(4d)}_{2+\ell,\ell}(1-\zb,1-z) \Bigg|_{\mathrm{small}\,z} + \mathcal{O}(\epsilon^5).
\end{equation}
This sum is most easily found using the language of twist conformal blocks \cite{Alday:2016njk}.
The individual conformal blocks have the particular form:
\begin{equation} \label{eq:4dTwist4BlockForm}
G^{(4d)}_{\tau+\ell,\ell}(1-\zbar,1-z) = \frac{(1-z)(1-\zbar)}{z-\zbar}\left(k_{\frac\tau2+\ell}(1-\zbar)k_{\frac\tau2-1}(1-z)-k_{\frac\tau2+\ell}(1-z)k_{\frac\tau2-1}(1-\zbar) \right),
\end{equation}
whence the small-\(z\) part of the sum factorizes.
Noticing that \( \gamma_{\RR,\ell}\sim 1/J_b^2\) to leading order, where \( J_b^2=\ell(\ell+1)\), one can subsequently relate the sums \( I_2^\RR \) to sums of the form
\begin{equation}
h^{(m)}_{\RR,2}(\zb) \equiv \sum_{\ell} \frac{a^{(0)}_{\RR,\ell}}{J_b^{2m}} k_{\ell+1}(1-\zb),
\end{equation}
with \(m=2\), where \(\ell=0,2,4,\ldots\) for the singlet and traceless symmetric representations, and \(\ell=1,3,5,\ldots\) for the antisymmetric representation.
 The \(a^{(0)}_{\RR,\ell}\) are the results for generalized free fields for \(\Delta_\varphi=1\), \(d=4\) in the appropriate representation:
\begin{equation}
a^{(0)}_{S,\ell}=\frac2N\frac{\Gamma(\ell+1)^2}{\Gamma(2\ell+1)},\quad 
a^{(0)}_{T,\ell}=\frac{2\,\Gamma(\ell+1)^2}{\Gamma(2\ell+1)},\quad 
a^{(0)}_{A,\ell}=-\frac{2\Gamma(\ell+1)^2}{\Gamma(2\ell+1)}.
\end{equation}
Note that the \( h^{(0)}_{\RR,2}(\zb)\) are precisely the sums that appears in the GFF result and are therefore already known.
The precise relation between the \(h^{(m)}\) is through a Casimir differential equation:
\begin{equation}
\bar{D}\,h^{(m+1)}_{\RR,2}(1-\zb) = h^{(m)}_{\RR,2}(1-\zb),
\end{equation}
where \(\bar{D} = (1-\zb)\zb^2\bar{\partial}^2 - \zb^2\bar{\partial} \), which has eigenvalue \(J_b^2\) on \(k_{\ell+1}(\zbar)\).
Solving these equations with the correct boundary conditions, we find the results for \(I_2^\RR \) listed in appendix \ref{app:I2andI4}.

\subsubsection{Quadrilinear operators}\label{sec:quadrilinears}
The discontinuities \( I_4^\RR\) arise because the Lagrangian contains a quartic coupling, so that new quadrilinear operators of approximate twist four appear.
These are formed out of four fundamental fields and derivatives and appear with (squared) OPE coefficients at order \( \epsilon^2 \).
At this order, no operators of approximate twists \( \tau = 6,8,\dots \) appear.
The twist four operators acquire anomalous dimensions at first order in \( \epsilon \), so that they contribute to the double discontinuity at order \( \cO(\epsilon^4)\).
For example, the singlet contribution takes the form
\begin{equation}\label{eq:I4preform}
I_4^S =  \frac{1}{8} \log^2(1-\zb) \sum_\ell \degsum{ a_{S,4,\ell} (\gamma_{S,4,\ell})^2 }G^{(4d)}_{4,\ell}(1-\zbar,1-z)+\cO(\epsilon^5).
\end{equation}
These operators are degenerate for spins \( \ell \geqslant 2 \) and we therefore use the notation \( \degsum{\,\cdot\,} \) for a sum over degenerate states:
\begin{equation}
\degsum{ a_{S,4,\ell} (\gamma_{S,4,\ell})^2 } \equiv \sum_i a_{S,4,\ell,i} (\gamma_{S,4,\ell,i})^2.
\end{equation}
If there was no problem with degenerate states, we could have computed first the OPE coefficients and anomalous dimensions using an order \(\epsilon^3\) inversion, and from this compute the sum \eqref{eq:I4preform}. 
We will employ this strategy later for the non-degenerate operators of spin~\(0\) and~\(1\). 
To compute the full \(I_4^\RR\), however, we will instead need to employ another strategy. We use the special form of the 4d conformal blocks, equation \eqref{eq:4dTwist4BlockForm}, and take the small-\(z\) limit to deduce that it must be of the following form 
\begin{equation} \label{eq:I4form}
I_4^\RR = \left( h_{\RR,4}(\zbar)\log z  - h_{\RR,4}(z) \log \zbar \right) \log^2(1-\zbar),\quad 
h_{\RR,4}(\zbar) = \sum_\ell \frac{1}{8} \degsum{a_{\RR,4,\ell} (\gamma_{\RR,4,\ell})^2 } k_{2+\ell}(\zbar).
\end{equation}
Guided by the structure of perturbation theory, and the results for double discontinuities to lower order, we make an ansatz for each \(h_{\RR,4}(\zbar)\) in terms of functions of pure transcendentality:
\begin{equation}
\left\{ 1, \,\log \zbar,\,\log^2 \zbar,\, \mathrm{Li}_2(1-\zbar),\,\log^3 \zbar, \,\log \zbar\ \mathrm{Li}_2(1-\zbar) ,\,\mathrm{Li}_3(1-\zbar),\,\mathrm{Li}_3\left(\frac{\zbar-1}{\zbar}\right)  \right\}.
\end{equation}
These form a basis of functions that are regular at \(\zbar\rightarrow1\) and are of pure transcendentality of maximal order three. 
We now fix the 24 constants in the ansatz.
Demanding, order by order in \( (1-\zbar)\), that \( h_{\RR,4} (\zbar) \) arises as the sum in equation \eqref{eq:I4form} leaves only nine undetermined constants in the total ansatz. 
Next, note that the \( I^\RR_4 \) are the only terms on the right-hand side of the crossing equations that give terms of the form \( \log^2 z\) and \( \log^3 z \). 
Since there are no \( \log^3 z \) terms on the left-hand side, this gives us three constraints on the constants in the ansatz.
The \(\log^2 z \) terms on the left-hand side are generated by sums of \( \ahat_{\RR,\ell}\big(\gamma^{(1)}_{\RR,\ell}\big)^2 \), which are known from our lower order computations, yielding three further constraints.
To see in more detail how this works, consider for example the crossing equation for the singlet representation, equation \eqref{eq:ONSingInv}.
The order \(\epsilon^2\) result for \(\gamma_{S,\ell}\) immmediately gives
\begin{equation}
U^{(2)}_{S,\hb} = \frac{ \left(2 g_S^2+g_T^2 (N-1)(N+2)\right)^2 \left(2 \bar{h}-1\right)}{2 N^3 (\bar{h}-1)^2 \bar{h}^2}+ \cO(\epsilon^5).
\end{equation}
The \( \log^2 z\) piece of this equation takes the form
\begin{equation}
\sum_\ell \frac{1}{8} \log^2 z ~ \ahat_{S,\ell} \left(\gamma_{S,\ell}\right)^2 z^{\Delta_\varphi} f_{\Delta,\ell}(\zbar)=  
\left. z^{\Delta_\varphi} \left(\frac{1}{N}I^S_4 + \frac{N^2+N-2}{2N^2}I^T_4 +\frac{1-N}{2N} I^A_4\right)\right|_{\mathrm{small}\,z,~ \log^2 z}
\end{equation}
On the right-hand side, each \( I^\RR_4 \) has a \(\log^2 z \) piece that can be readily computed in terms of the constants in the ansatz.
We then invert these terms and match to the known CFT-data \( \frac{1}{8}\ahat_{S,\ell} \big(\gamma_{S,\ell} \big)^2 \) on the left-hand side to gain an equation for the constants.
After doing this matching also for the \(T\) and \(A\) representations we are left with three undetermined constants.

\paragraph*{Fixing the constants using low spin operators.}
The last three constants in the ansatz will be fixed by using the fact that we have three non-degenerate quadrilinear operators, namely the singlet and traceless symmetric spin 0 operators, and the antisymmetric spin 1 operator.
As briefly mentioned above, we will first compute the OPE coefficients and anomalous dimensions of these operators by performing a projection of the order \(\epsilon^3\) inversion onto twist four operators. With \(a_{\RR,4,\ell} (\gamma_{\RR,4,\ell})^2\) at hand for the non-degenerate operators the last three constants in \(I_{4}^{\RR}\) are fixed by the decomposition \eqref{eq:I4form}.

Let us explain in more detail how to use the inversion formula to find the OPE coefficients and anomalous dimensions we need.
Recall that the conformal blocks in the small-\(z\) limit behave as \(z^{\tau/2}\). 
In the usual inversion problem, for leading twist operators, we can therefore focus on just the leading \(z\) dependence. 
When considering the inversion problem for operators with higher twists we will need to study also the subleading \(z\) dependence. 
We will therefore make use of a projection, defined below, and subleading expressions for the conformal blocks as derived in Appendix \ref{app:subleadBlocks}.
For simplicity we will focus on just the singlet representation as the method for the other representations is completely analogous.

On the right-hand side of the crossing equation, to cubic order, there are only the contributions from the identity operator \( \mathds{1}\) and the operators \( \varphi^2_S\) and \(\varphi^2_T\).
On the left-hand side, there is a `spillover' from the leading twist operators.
To compute this, we use the special form \eqref{eq:4dTwist4BlockForm} of the 4d blocks.
Since a double discontinuity can only arise from an infinite sum over the spin, we need only consider the piece
\begin{equation} \label{eq:4dBlockEnhPart}
G^{(4d), \mathrm{enh.}}_{\tau,\ell}(z,\zb) = -\frac{z\zb}{z-\zbar}k_{\frac\tau2-1}(z)k_{\frac\tau2+\ell}(\zb).
\end{equation}
With this in mind, to find the CFT-data of the leading twist operators we should have projected the crossing equation onto \( -\frac{z\zb}{z-\zbar}k_{\Delta_\varphi-1}(z)\). However this is equivalent to simply considering the small-\(z\) limit.
For the CFT-data of the quadrilinear operators, we do need to actually perform this projection onto \( -\frac{z\zb}{z-\zbar}k_{\Delta_\varphi}(z)\), which can be achieved by using the orthogonality relation
\begin{equation}\label{eq:kOrthogonality}
\frac{1}{2\pi \mathrm{i}} \oint \frac{\mathrm{d}{z}}{z^2} k_\alpha(z) k_{1-\beta} (z) = \delta_{\alpha\beta}.
\end{equation}
To order \( \epsilon^2\), the double discontinuity \( I_{\varphi^2}\) has a vanishing projection, while \( I_{\mathds{1}}\) simply gives the GFF contribution to the OPE coefficients.
However, on the left-hand side of the crossing equation there is a non-zero contribution from the anomalous dimensions of the leading twist operators.
In particular, the following sum has a non-zero projection onto \( -\frac{z\zb}{z-\zbar}k_{\Delta_\varphi}(z)\): 
\begin{equation} \label{eq:twist2SpilloverProk}
\sum_{\ell} a_{S,\ell}^{(0)} \gamma_{S,\ell} \partial_{\tau} G^{(4d), \mathrm{enh.}}_{\tau,\ell}(z,\zb)\bigg|_{\tau=2} 
\xrightarrow{\mathrm{projects}} -\frac{1}{4} \frac{z\zb}{z-\zb}k_1(z) \sum_{\ell} a_{S,\ell}^{(0)} \gamma_{S,\ell} k_{1+\ell}(\zb),
\end{equation}
where we used the 4d value \( \Delta_\varphi=1\) since the expression is already of order \(\epsilon^2\), and where the factor of \( \frac{1}{4}\) arises from the projection of
\begin{equation}
\partial_{\tau} k_{\frac{\tau}{2}-1} (z)\big|_{\tau=2} = \frac{1}{4} k_{1}(z) + \frac{1}{2}k_0(z) \log z\, .
\end{equation}

Implementing this we find that the correction to the OPE coefficients \( \degsum{a_{S,4,\ell}} = a_{S,4,\ell}^{GFF} + a_{S,4,\ell}^{(2)} + \cO(\epsilon^3) \) satisfies
\begin{equation}
-\frac{z\zb}{z-\zb}k_1(z) \sum_{\ell=0,2,\dots} \left( a_{S,4,\ell}^{(2)} k_{2+\ell}(\zb) + \frac{1}{4} a_{S,\ell}^{(0)} \gamma_{S,\ell} k_{1+\ell}(z) \right) \stackrel{\cdot}{=} 0\, ,
\end{equation}
where by \(\stackrel{\cdot}{=}\) we mean an equality of double discontinuities.
Note that both \( a_{S,4,\ell}^{GFF}\) and \(a_{S,4,\ell}^{(2)}\) are of order \(\cO(\epsilon^2)\), and we have ignored higher order terms.
From the kernel method \cite{Alday:2015ewa}, it follows that the double discontinuity of this sum can be computed by replacing the sum with an integral over \( (0,\infty)\).
Furthermore making a change \(\ell\rightarrow \ell +1\) in one of the terms gives the condition
\begin{equation}
\int_0^\infty \mathrm{d}\ell \left( a_{S,4,\ell}^{(2)} +\frac{1}{4} a_{S,\ell+1}^{(0)}\gamma_{S,\ell+1} \right) k_{2+\ell}(\zb) \stackrel{\cdot}{=} 0,
\end{equation}
which yields the result
\begin{equation}
a_{S,4,\ell}^{(2)} = -\frac{1}{4} a_{S,\ell+1}^{(0)} \gamma_{S,\ell+1} \, .
\end{equation}
By adding the GFF contribution, we then find the following results: 
\begin{align}
\degsum{a_{S,4,\ell} }&=\frac{\Gamma(\ell+2)^2}{\Gamma(2\ell+3)}\left(\frac{1}{N}\gamma_\varphi+\frac{2g_S^2+(N+2)(N-1)g_T^2}{4N^2(\ell+1)(\ell+2)}\right)+\cO(\epsilon^3), \label{eq:aS4}
\\
\degsum{a_{T,4,\ell} }&=  \frac{\Gamma(\ell+2)^2}{\Gamma(2\ell+3)}\left(\gamma_\varphi+\frac{2g_S^2+(N-2)g_T^2}{4N(\ell+1)(\ell+2)}  \right)+\cO(\epsilon^3),
\label{eq:aT4}
\\
\degsum{a_{A,4,\ell} }&=  \frac{\Gamma(\ell+2)^2}{\Gamma(2\ell+3)}\left(-\gamma_\varphi + \frac{-2g_S^2+(N+2)g_T^2}{4N(\ell+1)(\ell+2)} \right)+\cO(\epsilon^3),\label{eq:aA4}
\end{align}
where, as for the leading twist operators, the OPE coefficients of singlet and traceless operators vanish for odd spin, while those of the antisymmetric operators vanish for even spin. In these expressions \(\gamma_\varphi=\Delta_{\varphi}-\frac{d-2}2\) is the anomalous dimension of \(\varphi\), which starts at order \(\epsilon^2\).

To find the sums \(\degsum{a_{R,4,\ell}\gamma_{R,4,\ell}} \), we perform the same procedure at cubic order. 
In this case it is convenient to specialize to terms in the crossing equation that contain a factor of \( \log z\).
On the right-hand side, the double discontinuity \(I_{\varphi^2}\) contains such terms, while on the left-hand side there is a `spillover' contribution from the leading twist operators that arises as a dimensional correction to the conformal blocks:
\begin{equation}
\log z \sum_{\ell} a_{R,\ell}^{(0)} \frac{1}{2}\gamma_{R,\ell} \partial_d G^{(d)}_{2\Delta_\varphi,\ell}(z,\zb)\Big|_{d=4}.
\end{equation}
To find this contribution, we make use of the known form of the leading-\(z\) and subleading-\(z\) behaviour of the conformal blocks\footnote{We give some details of this computation in Appendix \ref{app:subleadBlocks}.}, which allows us to find the projections onto \( -\frac{z \zb}{z-\zb}k_{\Delta_\varphi}(z)\).
Inverting the right-hand side with the inversion formula, and making the appropriate shifts in the summation over \(\ell\) on the left-hand side, we then find the results (valid at order \(\epsilon^3\))
\begin{align}
\degsum{a_{S,4,\ell}\gamma_{S,4,\ell}} &= \frac{\Gamma (\ell+2)^2}{ (\ell+1) (\ell+2) \Gamma (2\ell+3)N}   \left( \frac{g_S^2 \left(\epsilon -g_S\right)}{N}+\frac{(N-1) (N+2) }{2N}g_T^2 \left(\epsilon -g_T\right)\right),\label{eq:aSgamma4}
\\
\degsum{a_{T,4,\ell}\gamma_{T,4,\ell}} &= \frac{\Gamma (\ell+2)^2}{ (\ell+1) (\ell+2) \Gamma (2\ell+3)N}   \left( g_S^2 \left(\epsilon -g_S\right)+\frac{N-2}{2} g_T^2 \left(\epsilon -g_T\right)\right) ,\label{eq:aTgamma4}
\\
\degsum{a_{A,4,\ell}\gamma_{A,4,\ell}} &= \frac{\Gamma (\ell+2)^2}{ (\ell+1) (\ell+2) \Gamma (2\ell+3)N}   \left(-g_S^2 \left(\epsilon -g_S\right)+\frac{N+2}{2} g_T^2 \left(\epsilon -g_T\right)\right).\label{eq:aAgamma4}
\end{align}
Combining equations \eqref{eq:aS4}--\eqref{eq:aA4} with \eqref{eq:aSgamma4}--\eqref{eq:aAgamma4}, the anomalous dimensions of the non-degenerate operators can be found, which then allows us to fix the final constants in the \( I_4^R\), which are given in Appendix \ref{app:I2andI4}.

As a check of this computation one can substitute the leading order expansions of \((g_S,g_T,\gamma_\varphi)\), which can be computed from the double discontinuities in section \ref{sec:uptocubicorder}, to find that
\begin{equation}
\gamma_{S,4,0} = \epsilon + \mathcal{O}(\epsilon^2),\qquad 
\gamma_{T,4,0} = \frac{8}{N+8}\epsilon + \mathcal{O}(\epsilon^2),\qquad 
\gamma_{A,4,1} = \frac{2}{N+8}\epsilon + \mathcal{O}(\epsilon^2),
\end{equation}
in perfect agreement with the literature \cite{Gliozzi:2017hni, Rychkov:2015naa,Liendo:2017wsn}\footnote{The result for the antisymmetric spin \(\ell =1\) operator was computed using the method described in \cite{Liendo:2017wsn}. We thank P. Liendo for making us aware of this method, which can be used to find the anomalous dimension of any operator to first order in the \(\epsilon\)-expansion.}.

\section{Constraints from low spins}\label{sec:lowLyingSpin}

As explained in the previous section, we have computed all double-discontinuities needed to find the CFT-data of leading twist operators to order \(\epsilon^4\).
Applying the inversion procedure thus gives expressions for \(U^{(p)}_{R,\hbar}\), and hence all CFT-data to order \(\epsilon^4\) in terms of \(\epsilon\) and the following parameters:
\begin{equation}\label{eq:allparameters}
 g_S,\quad g_T,\quad \gamma_\varphi,\quad \alpha_S,\quad \alpha_T.
\end{equation}
Notice that all these parameters depend perturbatively on \(\epsilon\). 
In this section we will see how all these constants can be determined in terms of \(\epsilon\), to leading and subleading order. 
This will be achieved by writing down consistency conditions for operators with low spin.

Firstly, there are two symmetry currents amongst the leading twist operators, namely the stress-energy tensor (\(S\), \(\ell=2\)) and the global symmetry current (\(A\), \(\ell=1\)). Their scaling dimensions are protected, and therefore we get
\begin{equation}
\Delta_{S,2}=d,\qquad\Delta_{A,1}=d-1,
\end{equation}
valid to all orders in perturbation theory. 
For example, the first of these equations gives at leading order \(\gamma_\varphi=\left(2g_S^2+(N+2)(N-1)g_T^2 \right)/(24N)+\cO(\epsilon^3)\). 
Using this in the second equation then gives the leading order relation \(g_T= \pm 2g_S/(N+2)+\cO(\epsilon^2)\).

Secondly, the assumption that the results for the singlet and traceless symmetric representations can be analytically continued all the way down to spin zero gives the following self-consistency conditions
\begin{equation}\label{eq:spinzeroconstraints}
\Delta_{S,0}=2\Delta_\varphi+g_S,\quad a_{S,0}=\frac2N\left(1+\alpha_S\right),\quad \Delta_{T,0}=2\Delta_\varphi+g_T,\quad a_{T,0}=2\left(1+\alpha_T\right).
\end{equation}
Here the left-hand sides of all these equations are computed from the \(U_{R,\hbar}^{(p)}\), and therefore depend on \(\epsilon\) and the constants in \eqref{eq:allparameters}.

The extrapolation to spin zero is subtle and works in the following way. Let us recall that the CFT-data was computed in the variable \(\hbar=\Delta_\varphi+\ell\), which is the most convenient variable for perturbative computations. However, the CFT-data is most naturally expressed in terms of the (full) conformal spin \(J_{\mathrm f}^2=\hbar_{\mathrm{f}}(\hbar_{\mathrm{f}}-1)\) \cite{Basso:2006nk}, conveniently written in terms of \(\hbar_{\mathrm{f}}=\hbar+\gamma_\ell/2=(\Delta+\ell)/2\).
Similar to the computation in \cite{Alday:2017zzv}, the expressions for the CFT-data are analytically continued past the pole \(\hbar_{\mathrm f}=1\), and evaluated for spin zero. For example, the equation for the leading correction to the singlet scalar operators becomes
\begin{equation}
\Delta_{S,0}=\left.2\Delta_{\varphi}-\frac{2g_S^2+(N+2)(N-1)g_T^2}{2N\hbar_{\mathrm f}(\hbar_{\mathrm f}-1)}\right|_{\hbar_{\mathrm f}=\Delta_\varphi+\frac{g_S}2}=2\Delta_{\varphi}+g_S,
\end{equation}
which is solved at order \(\epsilon\).\footnote{Notice that in this equation \(\hbar_{\mathrm{f}}=1+\cO(\epsilon)\), so that the denominator in the anomalous dimension is of order \(\epsilon\). \(\gamma_\varphi\) decouples from the equation at this order.} Together with the completely analogous equation for \(g_T\)  we get a coupled system of equations, which has two solutions,
\begin{equation}
g_S=0,\ g_T=0, \qquad \text {and }\qquad g_S=\frac{N+2}{N+8}\epsilon+\cO(\epsilon^2),\ g_T=\frac{2}{N+8}\epsilon+\cO(\epsilon^2).
\end{equation}
This result shows that, in the \(4-\epsilon\) expansion, any CFT with global \(\mathrm O(N)\) symmetry agrees to leading order with either the free theory or the critical \(\mathrm O(N)\) model! We emphasize that this result has followed purely from conformal and global symmetry with no input from Feynman diagrams.  
A similar equation has been derived in the large \(N\) expansion of \(\mathrm O(N)\) symmetric CFTs in \(2<d<4\) \cite{Alday:2016jfr}.

By using the full fourth order results for both anomalous dimensions and OPE coefficients, we can solve all four equations \eqref{eq:spinzeroconstraints} to subleading order. The results are
\begin{align}
g_S&=\frac{N+2}{N+8}\epsilon+\frac{6(N+2)(N+3)}{(N+8)^3}\epsilon^2+\cO(\epsilon^3),\label{eq:gSineps}\\
g_T&=\frac{2}{N+8}\epsilon+\frac{36+4N-N^2}{(N+8)^3}\epsilon^2+\cO(\epsilon^3),\label{eq:gTineps}\\
\gamma_{\varphi}&=\frac{N+2}{4
   (N+8)^2}\epsilon^2+\frac{(N+2) (272+56N-N^2) }{16 (N+8)^4}\epsilon ^3+\cO(\epsilon^4),
\\
\alpha_S&=-g_S-\frac{3}{N+2}g_S^2+\cO(\epsilon^3),\\ 
\alpha_T&=-g_T-\frac{N+6}{4}g_T^2+\cO(\epsilon^3),\label{eq:alphaTresult}
\end{align}
all in agreement with known results for the critical \(\mathrm O(N)\) model. This means that we can write down the complete results for the CFT-data, given in Appendix~\ref{app:CFTdata}. 
We have also checked that the results agree with known results in the large \(N\) expansion \cite{Lang:1993ge,Derkachov:1997ch}.


\section{Central charges and Pad\'e approximants}
\label{sec:CentralCharges}

From the results in the previous section we can extract the values to order \(\epsilon^4\) of the central charges \(C_T\) and \(C_J\). 
They are defined by the intrinsic normalization of the stress-energy tensor (\(S\), \(\ell=2\)) and the global symmetry current (\(A\), \(\ell=1\)) respectively and can be computed from their respective OPE coefficients using conformal Ward identities. 
Compared to the free theory in \(d=4-\epsilon\) dimensions they take the values\footnote{To facilitate a comparison with results in the literature, we have substituted the results of the previous section, as well as the expressions \eqref{eq:gS3}--\eqref{eq:gT3} for \(g_S\) and \(g_T\) at order \(\epsilon^3\).} 
\begin{align} \label{eq:CentralCharge}
\frac{C_T}{C_{T,\mathrm{free}}} &= 1
-\frac{5 (N+2) }{12 (N+8)^2} \epsilon ^2 
-\frac{(N+2) \left(7 N^2+382 N+1708\right)}{36 (N+8)^4} \epsilon ^3
\nonumber\\
&\quad-\frac{(N+2) \left(65 N^4+5998 N^3+309036 N^2+2396800N+5440832\right)}{1728 (N+8)^6}  \epsilon^4
\nonumber\\
&\quad+\frac{(N+2) \left(2 N^3+43 N^2+922 N+3488\right)\zeta_3}{12 (N+8)^5}  \epsilon^4+\cO(\epsilon^5),
\end{align}
and
\begin{align}\label{eq:JCharge}
\frac{C_J}{C_{J,\mathrm{free}}}&=1
-\frac{3 (N+2)}{4 (N+8)^2} \epsilon^2
-\frac{(N+2) \left(N^2+132 N+632\right) }{8 (N+8)^4}\epsilon^3
\nonumber\\
&\quad + \frac{(N+2) \left(11 N^4+246 N^3-13124 N^2-126976 N-310976\right)}{64 (N+8)^6} \epsilon ^4
\nonumber\\
&\quad +\frac{(N+2) \left(7 N^2+442 N+1792\right)\zeta_3}{4 (N+8)^5}\epsilon ^4+\cO(\epsilon^5).
\end{align}

The expressions up to order \( \epsilon^3\) are in perfect agreement with the ones computed in \cite{Dey:2016mcs}, whereas the \(\epsilon^4\) values are new results. 
For \(N=1\) there is no global symmetry current, and the central charge \(C_T\) result reduces to the one in \cite{Alday:2017zzv}.
Expanding the results \eqref{eq:CentralCharge} and \eqref{eq:JCharge} for large \(N\) we also get agreement with the \(1/N\) term in the \(\mathrm{O}(N)\) model given in \cite{Petkou:1995vu}. 

With the \(\epsilon^4\) results at hand we can compute some approximations for the value of the central charges in the strongly coupled theories in 3d.
A direct approach is to consider the expressions \eqref{eq:CentralCharge} and \eqref{eq:JCharge} evaluated at \(\epsilon=1\).
Including the fourth order improves the \(\epsilon^3\)-truncation results in \cite{Dey:2016mcs}, as can be seen in the first two columns of tables~\ref{tab:padeCT} and~\ref{tab:padeCJ}.

A more refined method to give approximate results in three dimensions is to use Pad\'e approximants. 
To construct a Pad\'e approximant for the central charge in the interval \(2<d<4\), we define
\begin{equation}\label{eq:cT2pluseps}
\mathrm{Pad\acute e}_{[m,n]}(d)=\frac{P_0+P_1d+P_2d^2+\ldots+P_md^m}{1+Q_1d+Q_2d^2+\ldots+Q_nd^n}.
\end{equation}
and subsequently fix the \(m+n+1\) constants such that \(\mathrm{Pad\acute e}_{[m,n]}(d)\) smoothly interpolates between the known expansions at \(d=4-\epsilon\) and at \(d=2+\epsilon\). 
Our results for the central charges allow the construction of Pad\'e approximants of one order higher than previously available in the literature.

\begin{table}
\centering

\begin{tabular}{|l||l|l|l|l|l|l|}
\hline
\(N\)&\(\epsilon^3\) trunc.&\(\epsilon^4\) trunc.&\(\mathrm{Pad\acute e}_{[4,2]}\)&\(\mathrm{Pad\acute e}_{[4,3]}\)&\(\mathrm{Pad\acute e}_{[5,2]}\)&Numeric bootstrap
\\\hline
 1& 0.957933 & 0.953972 & 0.948725* & 0.951076* & 0.951751* & 0.946600(22) \\
 2& 0.955556 & 0.951008 & 0.946610* & 0.948404* & 0.948791* & 0.94365(13) \\
 3& 0.955111 & 0.950485 & 0.933861 & 0.939515 & 0.946438 & 0.94418(43) \\
 4& 0.955729 & 0.951342 & 0.938871 & 0.943453 & 0.948923 & 0.94581(71) \\
 5 & 0.956919 & 0.952943 & 0.944714 & 0.948313 & 0.952052 & 0.9520(40) \\
 6& 0.958397 & 0.954911 & 0.949952 & 0.952772 & 0.955282 & 0.9547(41) \\
 10 & 0.964792 & 0.963212 & 0.964425 & 0.965578 & 0.966169 & 0.96394 \\
 20& 0.976230 & 0.977070 & 0.979761 & 0.980002 & 0.980052 & 0.97936 \\
 \hline
\end{tabular}

\caption{
Values for \({C_T}/{C_{T,\mathrm{free}}}\) in three dimensions using different methods of computation. 
The numeric values are taken from \cite{Kos:2013tga}.
*The \(N=1,2\) values denote the \(\mathrm{Pad\acute e}_{[3,1]}\), \(\mathrm{Pad\acute e}_{[3,2]}\) and \(\mathrm{Pad\acute e}_{[4,1]}\) respectively.
}\label{tab:padeCT}
\end{table}

Let us start by discussing the central charge \(C_T\) associated to the normalization of the stress-energy tensor.
The expression \eqref{eq:CentralCharge} in \(d=4-\epsilon\) is valid for all \(N\).
For \(N>2\), the critical \(\mathrm O(N)\) model is perturbatively described by the UV fixed-point of the non-linear sigma model in \(d=2+\epsilon\), and the central charge has the expansion \cite{Diab:2016spb}
\begin{equation}
\left.\frac{C_T}{C_{T,\mathrm{free}}}\right|_{d=2+\epsilon}=1-\frac1N+\frac{3(N-1)}{4N(N-2)}\epsilon^2+\cO(\epsilon^3).
\end{equation}
For \(N=1,2\), we use the values for the two-dimensional Ising model and the XY model respectively. 
In both these cases we have \(C_T/C_{T,\mathrm{free}}=1/2\) \cite{diFrancesco:1987aa}. 
We are not aware of any perturbative expansions in \(2+\epsilon\) dimensions for these values of \(N\)\footnote{Some numeric estimates are done in \cite{El-Showk:2013nia} for the \(N=1\) case in non-integer \(2<d<4\), but there is no known analytic expression.}, and therefore the corresponding Pad\'e approximants will be of two orders lower.
In table~\ref{tab:padeCT} we present the value at \(d=3\) for different Pad\'e approximants. 
One can see that for \(3\leqslant N\leqslant 6\) both new Pad\'e approximants give a value of the central charge closer to the values from numeric bootstrap, than \(\mathrm{Pad\acute e}_{[4,2]}\) which is computed using only the \(\epsilon^3\) result in \(d=4-\epsilon\).

\begin{figure}
\centering
\includegraphics[width=0.9\textwidth]{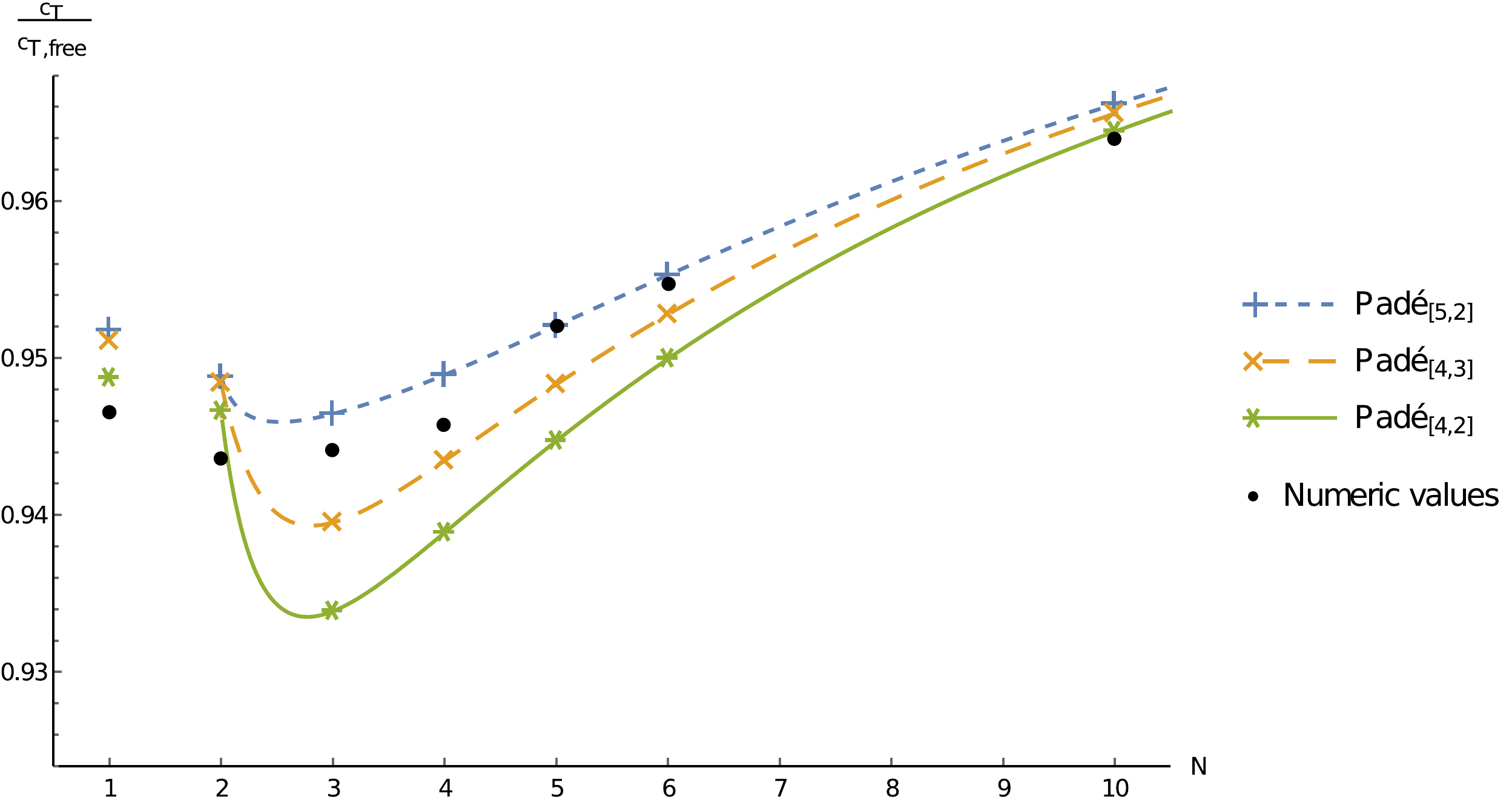}
\caption{Pad\'e approximants for \(C_T/C_{T,\mathrm{free}}\) in three dimensions.
For \(N=1,2\), the crosses show the values from the corresponding Pad\'e approximants of lower orders, as given in table~\ref{tab:padeCT}.
}\label{fig:padeCT}
\end{figure}

In figure~\ref{fig:padeCT} we show the results in three dimensions plotted against \(N\).
It is interesting to note that \(\left.\mathrm{Pad\acute e}_{[m,n]}(d)\right|_{N\rightarrow 2}=\left.\mathrm{Pad\acute e}_{[m-1,n-1]}(d)\right|_{N= 2}\), namely that the Pad\'e approximant constructed for \(N>2\) using \eqref{eq:cT2pluseps} reduces in the limit \(N\rightarrow2\) to the Pad\'e approximant for \(N=2\) of two orders lower.

\begin{table}
\centering
\begin{tabular}{|l||l|l|l|l|l|l|}
\hline
\(N\)&\(\epsilon^3\) trunc.&\(\epsilon^4\) trunc.&\(\mathrm{Pad\acute e}_{[3,2]}\)&\(\mathrm{Pad\acute e}_{[3,3]}\)&\(\mathrm{Pad\acute e}_{[4,2]}\)&Numeric bootstrap
\\\hline
 2 & 0.925 & 0.919049 & 
 &  &  &
0.9050(16)\\
 3 & 0.924740 & 0.919029 & 0.879552 & 0.896922 & 0.918319&0.9065(27) \\
 4 & 0.926215 & 0.921103 & 0.893426 & 0.906422 & 0.920658& \\
 5 & 0.928587 & 0.924237 & 0.905072 & 0.915022 & 0.924905& \\
 6& 0.931383 & 0.927839 & 0.914751 & 0.922546 & 0.929686& \\
 10 & 0.942901 & 0.942106 & 0.940299 & 0.943830 & 0.946478& \\
 20 & 0.962525 & 0.964566 & 0.966540 & 0.967555 & 0.968256&0.9674(8) \\
 \hline
\end{tabular}
\caption{
Values for \({C_J}/{C_{J,\mathrm{free}}}\) in three dimensions using different methods of computation. Numeric values are only available in the literature for \(N=2,3,20\) \cite{Kos:2015mba}.
}
\label{tab:padeCJ}
\end{table}

Finally we also construct Pad\'e approximants for \(C_J/C_{J,\mathrm{free}}\), whose values in three dimensions are shown in table~\ref{tab:padeCJ} together with the truncated \(4-\epsilon\) expansions. 
To construct these, we have used the expansion \cite{Diab:2016spb}
\begin{equation}\label{eq:cJ2pluseps}
\left.\frac{C_J}{C_{J,\mathrm{free}}}\right|_{d=2+\epsilon}=\frac{N-2}N+\frac1N\epsilon+\cO(\epsilon^2)
\end{equation}
from the non-linear sigma model for \(N\geqslant3\).


\section{Conclusion}
In this paper we used large spin perturbation theory to find the full CFT-data of the leading twist operators in the \(\mathrm{O}(N)\) model, to fourth order in the \( \epsilon\)-expansion.
An advantage of our approach is that the OPE coefficients, and the associated central charge corrections, are as easy to compute as anomalous dimensions, while they are generally significantly harder to compute using other methods.

Our derivation relies crucially on the inversion integrals from \cite{Caron-Huot:2017vep}, which are guaranteed to converge for spin \(\ell>1\).
However, by assuming that the expression the CFT-data can be continued to spin zero by the re-expanding in the full conformal spin, we found that all CFT-data agrees with that of either the free \(\mathrm O(N)\) theory or the critical \(\mathrm O(N)\) model.
It would be nice to gain a better understanding of precisely when such a continuation is possible.

We also gave a new result for the central charge corrections.
Using Pad{\' e} approximants, we provided predictions for these corrections in the strongly coupled 3d models discussed in the introduction, and our predictions compare favourably with previous predictions and with the known numerical results.

Finally, all our results are found to be compatible with known \( 1/N \) expansions.
It would be interesting to see if the expansions in terms of \(\epsilon\) and \(1/N\) also commute at the higher orders than currently available.
A bootstrap computation of CFT-data in the large \(N\) expansion would be of particular interest, since the complexity of a diagrammatic approach to large \(N\) increases very rapidly with the order of expansion.


\section*{Acknowledgements}
We would like to thank P. Liendo, D. Poland, D. Simmons-Duffin and in particular L.F. Alday for useful discussions.
M. van Loon is supported by an EPSRC studentship.


\appendix

\section{Conformal blocks in the small-\(z\) limit}
\label{app:subleadBlocks}
The conformal blocks \(G_{\tau,\ell}^{(d)}(z,\zb)\) are eigenfunctions of the shifted quadratic Casimir operator \(\mathcal{C}_{\tau,d}\) with eigenvalues the conformal spin \(J^2_{\tau,\ell} = \left(\frac{\tau}{2}+\ell-1\right)\left(\frac{\tau}{2}+\ell\right)\) \cite{Dolan:2003hv}. 
Since the Casimir \(\mathcal{C}_{\tau,d}\) respects orders in \(z\), i.e. 
\begin{equation}
\mathcal{C}_{\tau,d} \left( z^\alpha f_0(\zb) + \mathcal{O}(z^{\alpha+1}) \right) = z^\alpha \overline{\mathcal{C}}_{\tau,d,\alpha} \left(f_0(\zb)\right) + \mathcal{O}(z^{\alpha+1})
\end{equation}
for some reduced differential operator \(\overline{\mathcal{C}}_{\tau,d,\alpha}\), the Casimir equation can be solved order by order in \(z\).

Expanding the conformal block as
\begin{equation}
G_{\tau,\ell}^{(d)}(z,\zb) = z^{\frac{\tau}{2}} \sum_{k=0}^\infty z^k f^{(d)}_{k,\tau,\ell}(\zb),
\end{equation}
we find that the leading-\(z\) behaviour, given by \(f_{0,\tau,d,\ell}(\zb)\), solves a hypergeometric equation, with solution
\begin{equation}
f_{0,\tau,d,\ell}(\zb) = k_{\frac{\tau}{2}+\ell}(\zb) \equiv \zb^{\frac{\tau}{2}+\ell} \, _2 F_1 \left(\frac{\tau}{2}+\ell,\frac{\tau}{2}+\ell;\tau+2\ell;\zb \right),
\end{equation}
reproducing the well-known \(\mathrm{SL}(2,\mathbb{R}) \) conformal blocks.\footnote{The hypergeometric equation has two solutions, but we discard the unphysical solution that blows up as \(\zbar\rightarrow 0\).}

Inspired by \cite{Alday:2015ewa}, we make the following ansatz for the functions \(f_{n,\tau,\ell}^{(d)}\):
\begin{equation}
f^{(d)}_{n,\tau,\ell}(\zb) = \sum_{m=-n}^n c_{n,m}(d,\tau,\ell) k_{\frac{\tau}{2}+\ell+m}(\zb).
\end{equation}
Considering simultaneously the leading and subleading terms in the Casimir equation, we solve for the \(n=1\) coefficients
\begin{equation}
c_{1,-1} = \frac{(d-2)\ell}{d-4+2 \ell}, \qquad
c_{1,0} = \frac{\tau}{4}, \qquad
c_{1,1} = \frac{(d-2) (\ell+\tau -1) (2\ell+\tau )^2}{16 (2 \ell+\tau -1) (2 \ell+\tau +1) (2 (\ell+\tau +1)-d)}.
\end{equation}
This allows us to find in section \ref{sec:fourthorder} the dimensional corrections \( \partial_d G^{(d)}_{\tau,\ell}(z,\zb)\) to subleading order in \(z\). Solving the Casimir equation for higher values of \(n\) we find increasingly complicated functions \(c_{n,m}(\tau,d,\ell)\), which will not be needed in this paper.


\section{Results for \( I_2^\RR\) and \(I_4^\RR\)}
\label{app:I2andI4}
The results for the \(I_2^\RR\) are as follows:
\begin{align}
I_2^S \bigg|_{\mathrm{small}\,z} &= \frac{\log ^2\left(1-\bar{z}\right) \left((N+2)(N-1) g_T^2+2 g_S^2\right)^2 }{192 N^3}\Bigg[6 \mathrm{Li}_3\left(1-\bar{z}\right)-6
   \mathrm{Li}_3\left(\frac{\bar{z}-1}{\bar{z}}\right)\nonumber \\
 &\qquad\qquad\qquad +\log ^3\left(\bar{z}\right)+12 \log \left(\bar{z}\right)-6 \zeta _3+6 \zeta _2 \log (z)-12 \log (z)\Bigg]+ \mathcal{O}(\epsilon^5),\\
I_2^T \bigg|_{\mathrm{small}\,z} &= \frac{\log ^2\left(1-\bar{z}\right) \left((N-2) g_T^2+2 g_S^2\right)^2 }{192 N^2}\Bigg[6 \mathrm{Li}_3\left(1-\bar{z}\right)-6
   \mathrm{Li}_3\left(\frac{\bar{z}-1}{\bar{z}}\right)\nonumber \\
 &\qquad\qquad\qquad +\log ^3\left(\bar{z}\right)+12 \log \left(\bar{z}\right)-6 \zeta _3+6 \zeta _2 \log (z)-12 \log (z)\Bigg]+ \mathcal{O}(\epsilon^5),\\
I_2^A \bigg|_{\mathrm{small}\,z} &= \frac{\log ^2\left(1-\bar{z}\right) \left((N+2) g_T^2-2 g_S^2\right)^2}{64 N^2} \Bigg[-6 \mathrm{Li}_3\left(1-\bar{z}\right)-6 \mathrm{Li}_3\left(\frac{\bar{z}-1}{\bar{z}}\right)\nonumber\\
 &\qquad\qquad\qquad+4 \mathrm{Li}_2\left(1-\bar{z}\right) \log \left(\bar{z}\right)+\log ^3\left(\bar{z}\right)+6 \zeta _3+2 \zeta _2 \log (z)\Bigg]+ \mathcal{O}(\epsilon^5).
\end{align}

We do not give the results for the \(I_4^\RR\) directly, but give the functions \( h_{\RR,4}(\zbar) \) defined in \eqref{eq:I4form}:
\begin{align}
h_{S,4}(\zb) &= \frac{1}{4} \Gamma _{S,0} \bigg[-6 \mathrm{Li}_3\left(1-\zb\right)-6 \mathrm{Li}_3\left(\frac{\zb-1}{\zb}\right)+4 \mathrm{Li}_2\left(1-\zb\right) \log
   \left(\zb\right)+\log ^3\left(\zb\right)\bigg]  \nonumber\\
   & \quad   -\frac{(N+2)(N-1) g_T^4+2 g_S^4}{32N^2}   \bigg[-12 \mathrm{Li}_3\left(1-\zb\right)-12 \mathrm{Li}_3\left(\frac{\zb-1}{\zb}\right)\nonumber\\
   & \qquad\qquad\qquad   +8\mathrm{Li}_2\left(1-\zb\right) \log \left(\zb\right)+2 \log ^3\left(\zb\right)-\log ^2\left(\zb\right)\bigg] + \mathcal{O}(\epsilon^5) , \\
h_{T,4}(\zb) &= \frac{1}{4} \Gamma _{T,0} \bigg[-6 \mathrm{Li}_3\left(1-\zb\right)-6 \mathrm{Li}_3\left(\frac{\zb-1}{\zb}\right)+4 \mathrm{Li}_2\left(1-\zb\right) \log
   \left(\zb\right)+\log ^3\left(\zb\right)\bigg]\nonumber \\
 &  \quad -\frac{ \left((N-2) (N+4) g_T^2+8 g_S^2\right)g_T^2}{64N} \bigg[-12 \mathrm{Li}_3\left(1-\zb\right)-12 \mathrm{Li}_3\left(\frac{\zb-1}{\zb}\right)\nonumber \\
 &\qquad \qquad\qquad +8 \mathrm{Li}_2\left(1-\zb\right) \log \left(\zb\right)+2 \log ^3\left(\zb\right)-\log ^2\left(\zb\right)\bigg] + \mathcal{O}(\epsilon^5), \\
h_{A,4}(\zb) &= -\frac{9}{4} \Gamma _{A,1} \bigg[6 \mathrm{Li}_3\left(1-\zb\right)-6 \mathrm{Li}_3\left(\frac{\zb-1}{\zb}\right)+\log ^3\left(\zb\right)+12 \log \left(\zb\right)\bigg] \nonumber\\
&\quad  -\frac{1}{64} (N+2) g_T^4 \bigg[4 \mathrm{Li}_2\left(1-\zb\right)+12 \mathrm{Li}_3\left(1-\zb\right)-12 \mathrm{Li}_3\left(\frac{\zb-1}{\zb}\right) +2 \log ^3\left(\zb\right)   \nonumber \\
&\qquad\qquad\qquad  +\log ^2\left(\zb\right) +28 \log \left(\zb\right)\bigg]+ \mathcal{O}(\epsilon^5) ,
\end{align}
where we defined \( \Gamma_{R,\ell} =a_{R,4,\ell} \gamma_{R,4,\ell}^2 \), which were computed in section~\ref{sec:quadrilinears} with the following result:
\begin{align}
\Gamma_{S,0} &= \frac{\left((N-1) (N+2) g_T^2 \left(\epsilon -g_T\right)+2 \epsilon  g_S^2-2 g_S^3\right)^2}{4 N^2 \left((N-1) (N+2) g_T^2+2 g_S^2+8 N \gamma _{\varphi }\right)} + \mathcal{O}(\epsilon^5), \\
\Gamma_{T,0} &= \frac{\left((N-2) g_T^2 \left(\epsilon -g_T\right)+2 \epsilon  g_S^2-2 g_S^3\right)^2}{4 N \left((N-2) g_T^2+2 g_S^2+8 N \gamma _{\varphi }\right)}+ \mathcal{O}(\epsilon^5) , \\
\Gamma_{A,1} &= \frac{\left((N+2) g_T^2 \left(\epsilon -g_T\right)-2 \epsilon  g_S^2+2 g_S^3\right)^2}{36 N \left((N+2) g_T^2-2 g_S^2-24 N \gamma _{\varphi }\right)}+ \mathcal{O}(\epsilon^5) .
\end{align}


\section{Results for CFT-data}
\label{app:CFTdata}
In this appendix we present the results for the functions \(U^{(p)}_{\RR,\hbar}\),
which are related to the anomalous dimensions and OPE coefficients through equation \eqref{eq:DefnOfU}:
\begin{equation}
\ahat_{\RR,\Delta}(\hbar) \left( \gamma_{\RR,\ell} \right)^p = U^{(p)}_{\RR,\hbar} + \frac{1}{2} \partial_{\hbar} U^{(p+1)}_{\RR,\hbar}+ \frac{1}{8} \partial^2_{\hbar} U^{(p+2)}_{\RR,\hbar} + \dots ,
\end{equation}
where the \(\ahat(\hbar) \) are in turn defined by\footnote{Note that since our definition \eqref{eq:4dTwist4BlockForm} of the conformal blocks differ from \cite{Dolan:2000ut} by a factor \((-2)^{-\ell}\), the OPE coefficients for the antisymmetric representation are negative.}
\begin{equation}
 a_{\RR,\Delta,\ell} = r_{\frac{\Delta+\ell}2} \ahat_{\RR}(\hbar),
 \qquad \hbar=\Delta_\varphi+\ell,
 \qquad r_h = \frac{\Gamma(h)^2}{\Gamma(2h)}.
\end{equation}
The results for the \(U^{(p)}_{\RR,\hbar}\), as well as explicit results for the scaling dimensions and OPE coefficients, can be found in the Mathematica file included with this paper. In the functions \(U^{(p)}_{\RR,\hbar}\) below we have substituted the second order solutions for the parameters of the theory according to equations \eqref{eq:gSineps}--\eqref{eq:alphaTresult}. 
In order to compare with the literature, one needs to substitute also the \(\epsilon^3\) terms of \(g_S\) and \(g_T\), given by 
\cite{Brezin:1973aa,Braun:2013tva}\footnote{The fourth order correction to \(\Delta_\varphi\) also enters the results, but is fixed in terms of \(g_S\) and \(g_T\) using stress-energy tensor conservation at order \(\epsilon^4\). In the explicit expressions for the CFT-data in the Mathematica file, we have done this substitution.}
\begin{align}\label{eq:gS3}
g_S^{(3)}&=-\frac{(N+2) \left(N^3-202 N^2-976 N-1568\right) }{4 (N+8)^5}-\frac{12  (N+2) (5 N+22)}{(N+8)^4}\zeta _3,\\\label{eq:gT3}
g_T^{(3)}&=-\frac{N^4+45 N^3+190 N^2-144 N-1568 }{2 (N+8)^5}-\frac{24(5 N+22)}{(N+8)^4} \zeta _3,
\end{align}
where \(\zeta_k\) denote Riemann's zeta values.

In the equations below, the harmonic sums have argument \(S_a=S_a(\hbar-1)\). For \(S_{-2}\) we always use the analytic continuation from even argument, \(S_{-2}(x)=\frac14\left(\psi^{(1)}\left(\frac{x+1}2\right)-\psi^{(1)}\left(\frac{x+2}2\right)\right)-\frac12\zeta_2\). For ease of notation, we also write \( J^2 = \hb (\hb-1)\).

Results for \( U^{(2)}_{R,\hb}\):
\begin{align}
U^{(2)}_{S,\hb} &= \frac{18 (N+2)^2(2\hb-1)}{N (N+8)^4 J^4} \epsilon^4, \\
U^{(2)}_{T,\hb} &= \frac{2 (N+6)^2 (2 \hb-1)}{(N+8)^4 J^4} \epsilon ^4, \\
U^{(2)}_{A,\hb} &= -\frac{2 (N+2)^2 (2 \hb-1)}{(N+8)^4 J^4} \epsilon ^4.
\end{align}

Results for \( U^{(1)}_{R,\hb}\):
\begin{align}
&U^{(1)}_{S,\hb} = \frac{ (2 \hb-1) (N+2)}{N (N+8)^2 J^2}\Bigg[
-6 \epsilon ^2+\epsilon ^3\frac{2  \left(5 N^2+26 N+68+(N+8)^2 S_1\right)}{(N+8)^2}
+\epsilon ^4 \Bigg(\frac{9 (N+2)}{J^4 (N+8)^2}\nonumber\\
&\quad+\frac{16 (5 N+22) S_{-2}+21 (N+2)}{2 J^2 (N+8)^2} 
-\frac{ \left(N^2+6 N+20\right) S_1^2 + 2(N+2)^2 S_1+2 (5 N+22) S_{-2}}{(N+8)^2}
\nonumber\\
&\quad 
+
\frac{-6 N^4-51 N^3+132 N^2+2856 N+7464}{(N+8)^4}
-\frac{\zeta _2}{2} 
-\frac{4 (N+8) (g_S^{(3)}+(N-1) g_T^{(3)})}{N} 
\Bigg) \Bigg],\\
&U^{(1)}_{T,\hb} =
-\frac{2 \hb-1}{2 J^2 (N+8)^2} \Bigg[ 
4 (N+6) \epsilon ^2
-\epsilon^3 \Bigg(\frac{4 (N (N+6)+16) S_1}{N+8} 
\nonumber\\
&\quad + \frac{4 (N (N (N+12)+64)+136)}{(N+8)^2} \Bigg)
+\epsilon^4 \Bigg(-\frac{(N+6) (3 N+14)+32 (5 N+22) S_{-2}}{J^2 (N+8)^2}
\nonumber\\
&\quad -\frac{2 (N+6)^2}{J^4 (N+8)^2} 
+\frac{4(N (N+16)+44) S_{-2}-2(N^3+8N^2+24N+40)S_1^2}{(N+8)^2}
\nonumber\\
&\quad
-\frac{22 N^4+492 N^3+4512 N^2+18704 N+29856}{(N+8)^4}
+\frac{\zeta _2 (16-(N-2) N)}{N+8}
\nonumber\\
&\quad
+\frac{4 (N+2)^2 \left(N^2+16\right) S_1}{(N+8)^3}
+\frac{8 (N+8) ((N+2) g_S^{(3)}+(N-2) g_T^{(3)})}{N}
\Bigg) \Bigg],\\
&U^{(1)}_{A,\hb} =
\frac{(2 \hb-1) (N+2)}{2 J^2 (N+8)^2} \Bigg[ 
4 \epsilon ^2
-4 \epsilon ^3 \frac{ (N+4) (N+8) S_1+N (N+2)+12}{(N+8)^2}
\nonumber\\
&\quad + \epsilon^4 \Bigg(
-\frac{2 (N+2)}{J^4 (N+8)^2} 
-\frac{3 (N+2)}{J^2 (N+8)^2}
-\frac{46 N^3+696 N^2+3984 N+7280}{(N+8)^4}
\nonumber\\
&\quad +\frac{4 (N-2) (N+2)^2 S_1}{(N+8)^3}
+\frac{2 (N (N+6)+12) S_1^2}{(N+8)^2}
+\frac{4 (N-2) S_{-2}}{(N+8)^2}
-\frac{\zeta _2 (N (N+4)+8)}{(N+8)^2}
\nonumber\\&\quad
+\frac{8 (N+8) (g_S^{(3)}-g_T^{(3)})}{N}
\Bigg) \Bigg].
\end{align}

Results for \( U^{(0)}_{R,\hb}\):
\begin{align}
&U^{(0)}_{S,\hb} = \frac{1}{N}\ahat^{GFF}(\hb) -\frac{(2 \hb-1) (N+2)}{4  N (N+8)^2J^2}
\Bigg[\frac{12 }{J^2} \epsilon ^2
-4 \epsilon ^3 \left(2 \zeta _2+\frac{N (5 N+26)+68}{J^2 (N+8)^2}+\frac{S_1}{J^2}\right)
\nonumber\\
& \quad + \epsilon^4 \Bigg(
-\frac{6 (N+2)}{J^4 (N+8)^2}
+\frac{12 (N (N (N (N+8)-31)-524)-1308)}{J^2 (N+8)^4}
+6 S_3
\nonumber\\
& \quad + \frac{\zeta _2 ((N-2) N+28)+4 (N+2)^2 S_1+2 (N (N+6)+20) S_1^2 + 4 (5 N+22) (6\zeta_3-S_{-2})}{J^2 (N+8)^2}
\nonumber\\
&\quad +\frac{2 \zeta _2 \left(4 (5 N+22) S_1+N (8 N+47)+134\right)}{(N+8)^2}
- \frac{6 \zeta _3 (N (N+21)+86)}{(N+8)^2}\nonumber\\
&\quad+\frac{8 (N+8) (g_S^{(3)}+(N-1) g_T^{(3)})}{J^2 N}
\Bigg) \Bigg],\\
&U^{(0)}_{T,\hb} = \ahat^{GFF}(\hb) + \frac{2 \hb-1}{4 J^2 (N+8)^2} \Bigg[ -\frac{4 (N+6)}{J^2} \epsilon^2 + \epsilon^3 \Bigg( \frac{4 (N (N (N+12)+64)+136)}{J^2 (N+8)^2} 
\nonumber\\
&\quad + \frac{4 (N (N+6)+16) S_1}{J^2 (N+8)} + \frac{32 \zeta _2 (N+4)}{N+8} \Bigg)
+\epsilon^4\Bigg( \frac{2 (N+2) (N+6)}{J^4 (N+8)^2}+ \frac{N^2 \left(\zeta _2-2 S_1^2\right)}{J^2 (N+8)}
\nonumber\\
&\quad 
+\frac{12(N(N(N(2N+45)+410)+1676)+2616)}{J^2 (N+8)^4}
-\frac{8 \zeta _2 (N (3 N+20)+44) S_1}{(N+8)^2} 
\nonumber\\
&\quad -\frac{4 (N+2)^2 \left(N^2+16\right) S_1}{J^2 (N+8)^3}
 +\frac{2 \zeta _3 (N (N (N+25)+208)+516)}{(N+8)^2}
 -2 (N+6) S_3
\nonumber\\
&\quad +\frac{4 \left((N (N+16)+44) S_{-2}-2 \zeta _2 (N+7)-4 (3 N+5) S_1^2\right)}{J^2 (N+8)^2}
-\frac{48\zeta_3(5N+22)}{J^2(N+8)^2}
\nonumber\\
&\quad 
-\frac{2 \zeta _2 (N (N (33 N+296)+1244)+2144)}{(N+8)^3}
-\frac{8 (N+8) ((N+2) g_S^{(3)}+(N-2) g_T^{(3)})}{J^2 N}
\Bigg) \Bigg],\\
&U^{(0)}_{A,\hb} = -\ahat^{GFF}(\hb) 
+ \frac{(2 \hb-1) (N+2)}{4 J^2 (N+8)^2}\Bigg[
\frac{4}{J^2} \epsilon ^2
-\epsilon^3 \Bigg(
\frac{4 (N (N+2)+12)}{J^2 (N+8)^2} +\frac{4 (N+4) S_1}{J^2 (N+8)}
\nonumber\\
&\quad +\frac{16 \zeta _2}{N+8} \Bigg) 
+\epsilon^4\Bigg( 
\frac{-2 (N+2)}{J^4 (N+8)^2}
-\frac{12 \left(4 N^3+61 N^2+348 N+628\right)}{J^2 (N+8)^4}
+\frac{8 \zeta _2 (3 N+10) S_1}{(N+8)^2}
\nonumber\\
&\quad +\frac{4 (N-2) (N+2)^2 S_1}{J^2 (N+8)^3}
+\frac{(N (N+6)+12) \left(2 S_1^2-\zeta _2\right)+4 (N-2) S_{-2}}{J^2 (N+8)^2}+2 S_3
\nonumber\\
&\quad +\frac{2 \zeta _2 (N (17 N+74)+224)}{(N+8)^3}
-\frac{2 \zeta _3 (N (N+13)+70)}{(N+8)^2}
+
\frac{8 (N+8) (g_S^{(3)}-g_T^{(3)})}{J^2 N}
\Bigg) \Bigg].
\end{align}
In these expressions
\begin{equation}
\ahat^{GFF}(\hb) = \frac{2(2\hbar-1)\Gamma(\hbar+\Delta_\varphi-1)}{\Gamma(\Delta_\varphi)^2\Gamma(\hbar-\Delta_\varphi+1)}.
\end{equation}

\bibliographystyle{JHEP}
\bibliography{ONmodel}

\providecommand{\href}[2]{#2}\begingroup\raggedright\begin{thebibliography}{10}

\bibitem{Rattazzi:2008pe}
R.~Rattazzi, V.~S. Rychkov, E.~Tonni, and A.~Vichi, {\it {Bounding scalar
  operator dimensions in 4D CFT}},  {\em JHEP} {\bf 12} (2008) 031,
  [\href{http://arxiv.org/abs/0807.0004}{{\tt arXiv:0807.0004}}].

\bibitem{Alday:2007mf}
L.~F. Alday and J.~M. Maldacena, {\it {Comments on operators with large spin}},
   {\em JHEP} {\bf 11} (2007) 019, [\href{http://arxiv.org/abs/0708.0672}{{\tt
  arXiv:0708.0672}}].

\bibitem{Fitzpatrick:2012yx}
A.~L. Fitzpatrick, J.~Kaplan, D.~Poland, and D.~Simmons-Duffin, {\it {The
  Analytic Bootstrap and AdS Superhorizon Locality}},  {\em JHEP} {\bf 12}
  (2013) 004, [\href{http://arxiv.org/abs/1212.3616}{{\tt arXiv:1212.3616}}].

\bibitem{Komargodski:2012ek}
Z.~Komargodski and A.~Zhiboedov, {\it {Convexity and Liberation at Large
  Spin}},  {\em JHEP} {\bf 11} (2013) 140,
  [\href{http://arxiv.org/abs/1212.4103}{{\tt arXiv:1212.4103}}].

\bibitem{Alday:2015eya}
L.~F. Alday, A.~Bissi, and T.~\L{}ukowski, {\it {Large spin systematics in
  CFT}},  {\em JHEP} {\bf 11} (2015) 101,
  [\href{http://arxiv.org/abs/1502.07707}{{\tt arXiv:1502.07707}}].

\bibitem{Alday:2016njk}
L.~F. Alday, {\it {Large Spin Perturbation Theory for Conformal Field
  Theories}},  {\em Phys. Rev. Lett.} {\bf 119} (2017), no.~11 111601,
  [\href{http://arxiv.org/abs/1611.01500}{{\tt arXiv:1611.01500}}].

\bibitem{Caron-Huot:2017vep}
S.~Caron-Huot, {\it {Analyticity in Spin in Conformal Theories}},  {\em JHEP}
  {\bf 09} (2017) 078, [\href{http://arxiv.org/abs/1703.00278}{{\tt
  arXiv:1703.00278}}].

\bibitem{Alday:2017vkk}
L.~F. Alday and S.~Caron-Huot, {\it {Gravitational S-matrix from CFT dispersion
  relations}},  \href{http://arxiv.org/abs/1711.02031}{{\tt arXiv:1711.02031}}.

\bibitem{Simmons-Duffin:2017nub}
D.~Simmons-Duffin, D.~Stanford, and E.~Witten, {\it {A spacetime derivation of
  the Lorentzian OPE inversion formula}},  {\em JHEP} {\bf 07} (2018) 085,
  [\href{http://arxiv.org/abs/1711.03816}{{\tt arXiv:1711.03816}}].

\bibitem{Alday:2016jfr}
L.~F. Alday, {\it {Solving CFTs with Weakly Broken Higher Spin Symmetry}},
  {\em JHEP} {\bf 10} (2017) 161, [\href{http://arxiv.org/abs/1612.00696}{{\tt
  arXiv:1612.00696}}].

\bibitem{Alday:2017xua}
L.~F. Alday and A.~Bissi, {\it {Loop Corrections to Supergravity on $AdS_5
  \times S^5$}},  {\em Phys. Rev. Lett.} {\bf 119} (2017), no.~17 171601,
  [\href{http://arxiv.org/abs/1706.02388}{{\tt arXiv:1706.02388}}].

\bibitem{Dey:2017fab}
P.~Dey, K.~Ghosh, and A.~Sinha, {\it {Simplifying large spin bootstrap in
  Mellin space}},  {\em JHEP} {\bf 01} (2018) 152,
  [\href{http://arxiv.org/abs/1709.06110}{{\tt arXiv:1709.06110}}].

\bibitem{Henriksson:2017eej}
J.~Henriksson and T.~\L{}ukowski, {\it {Perturbative Four-Point Functions from
  the Analytic Conformal Bootstrap}},  {\em JHEP} {\bf 02} (2018) 123,
  [\href{http://arxiv.org/abs/1710.06242}{{\tt arXiv:1710.06242}}].

\bibitem{vanLoon:2017xlq}
M.~van Loon, {\it {The Analytic Bootstrap in Fermionic CFTs}},  {\em JHEP} {\bf
  01} (2018) 104, [\href{http://arxiv.org/abs/1711.02099}{{\tt
  arXiv:1711.02099}}].

\bibitem{Alday:2017zzv}
L.~F. Alday, J.~Henriksson, and M.~van Loon, {\it {Taming the
  $\epsilon$-expansion with large spin perturbation theory}},  {\em JHEP} {\bf
  07} (2018) 131, [\href{http://arxiv.org/abs/1712.02314}{{\tt
  arXiv:1712.02314}}].

\bibitem{Pelissetto:2000ek}
A.~Pelissetto and E.~Vicari, {\it {Critical phenomena and renormalization group
  theory}},  {\em Phys. Rept.} {\bf 368} (2002) 549--727,
  [\href{http://arxiv.org/abs/cond-mat/0012164}{{\tt cond-mat/0012164}}].

\bibitem{Derkachov:1997pf}
S.~E. Derkachov, J.~A. Gracey, and A.~N. Manashov, {\it {Four loop anomalous
  dimensions of gradient operators in \(\phi^4\) theory}},  {\em Eur. Phys. J.}
  {\bf C2} (1998) 569--579, [\href{http://arxiv.org/abs/hep-ph/9705268}{{\tt
  hep-ph/9705268}}].

\bibitem{Manashov:2017xtt}
A.~N. Manashov, E.~D. Skvortsov, and M.~Strohmaier, {\it {Higher spin currents
  in the critical $O(N$) vector model at $1/N^{2}$}},  {\em JHEP} {\bf 08}
  (2017) 106, [\href{http://arxiv.org/abs/1706.09256}{{\tt arXiv:1706.09256}}].

\bibitem{Lang:1993ge}
K.~Lang and W.~Ruhl, {\it {Critical O(N) vector nonlinear sigma models: A
  Resume of their field structure}},  1993.
\newblock \href{http://arxiv.org/abs/hep-th/9311046}{{\tt hep-th/9311046}}.

\bibitem{Petkou:1995vu}
A.~C. Petkou, {\it {C(T) and C(J) up to next-to-leading order in \(1/N\) in the
  conformally invariant \(O(N)\) vector model for \(2 < d < 4\)}},  {\em Phys.
  Lett.} {\bf B359} (1995) 101--107,
  [\href{http://arxiv.org/abs/hep-th/9506116}{{\tt hep-th/9506116}}].

\bibitem{Derkachov:1997ch}
S.~E. Derkachov and A.~N. Manashov, {\it {The Simple scheme for the calculation
  of the anomalous dimensions of composite operators in the 1/N expansion}},
  {\em Nucl. Phys.} {\bf B522} (1998) 301--320,
  [\href{http://arxiv.org/abs/hep-th/9710015}{{\tt hep-th/9710015}}].

\bibitem{Diab:2016spb}
K.~Diab, L.~Fei, S.~Giombi, I.~R. Klebanov, and G.~Tarnopolsky, {\it {On
  ${C}_{J}$ and ${C}_{T}$ in the Gross–Neveu and O(N) models}},  {\em J.
  Phys.} {\bf A49} (2016), no.~40 405402,
  [\href{http://arxiv.org/abs/1601.07198}{{\tt arXiv:1601.07198}}].

\bibitem{Gliozzi:2017hni}
F.~Gliozzi, A.~L. Guerrieri, A.~C. Petkou, and C.~Wen, {\it {The analytic
  structure of conformal blocks and the generalized Wilson-Fisher fixed
  points}},  {\em JHEP} {\bf 04} (2017) 056,
  [\href{http://arxiv.org/abs/1702.03938}{{\tt arXiv:1702.03938}}].

\bibitem{Fitzpatrick:2011dm}
A.~L. Fitzpatrick and J.~Kaplan, {\it {Unitarity and the Holographic
  S-Matrix}},  {\em JHEP} {\bf 10} (2012) 032,
  [\href{http://arxiv.org/abs/1112.4845}{{\tt arXiv:1112.4845}}].

\bibitem{Dolan:2000ut}
F.~A. Dolan and H.~Osborn, {\it {Conformal four point functions and the
  operator product expansion}},  {\em Nucl. Phys.} {\bf B599} (2001) 459--496,
  [\href{http://arxiv.org/abs/hep-th/0011040}{{\tt hep-th/0011040}}].

\bibitem{Alday:2015ewa}
L.~F. Alday and A.~Zhiboedov, {\it {An Algebraic Approach to the Analytic
  Bootstrap}},  {\em JHEP} {\bf 04} (2017) 157,
  [\href{http://arxiv.org/abs/1510.08091}{{\tt arXiv:1510.08091}}].

\bibitem{Rychkov:2015naa}
S.~Rychkov and Z.~M. Tan, {\it {The $\epsilon$-expansion from conformal field
  theory}},  {\em J. Phys.} {\bf A48} (2015), no.~29 29FT01,
  [\href{http://arxiv.org/abs/1505.00963}{{\tt arXiv:1505.00963}}].

\bibitem{Liendo:2017wsn}
P.~Liendo, {\it {Revisiting the dilatation operator of the Wilson–Fisher
  fixed point}},  {\em Nucl. Phys.} {\bf B920} (2017) 368--384,
  [\href{http://arxiv.org/abs/1701.04830}{{\tt arXiv:1701.04830}}].

\bibitem{Basso:2006nk}
B.~Basso and G.~P. Korchemsky, {\it {Anomalous dimensions of high-spin
  operators beyond the leading order}},  {\em Nucl. Phys.} {\bf B775} (2007)
  1--30, [\href{http://arxiv.org/abs/hep-th/0612247}{{\tt hep-th/0612247}}].

\bibitem{Dey:2016mcs}
P.~Dey, A.~Kaviraj, and A.~Sinha, {\it {Mellin space bootstrap for global
  symmetry}},  {\em JHEP} {\bf 07} (2017) 019,
  [\href{http://arxiv.org/abs/1612.05032}{{\tt arXiv:1612.05032}}].

\bibitem{Kos:2013tga}
F.~Kos, D.~Poland, and D.~Simmons-Duffin, {\it {Bootstrapping the $O(N)$ vector
  models}},  {\em JHEP} {\bf 06} (2014) 091,
  [\href{http://arxiv.org/abs/1307.6856}{{\tt arXiv:1307.6856}}].

\bibitem{diFrancesco:1987aa}
P.~di~Francesco, H.~Saleur, and J.~B. Zuber, {\it {Relations between the
  Coulomb Gas Picture and Conformal Invariance of Two-Dimensional Critical
  Models}},  {\em J. Stat. Phys.} {\bf 49} (1987) 57.

\bibitem{El-Showk:2013nia}
S.~El-Showk, M.~Paulos, D.~Poland, S.~Rychkov, D.~Simmons-Duffin, and A.~Vichi,
  {\it {Conformal Field Theories in Fractional Dimensions}},  {\em Phys. Rev.
  Lett.} {\bf 112} (2014) 141601, [\href{http://arxiv.org/abs/1309.5089}{{\tt
  arXiv:1309.5089}}].

\bibitem{Kos:2015mba}
F.~Kos, D.~Poland, D.~Simmons-Duffin, and A.~Vichi, {\it {Bootstrapping the
  O(N) Archipelago}},  {\em JHEP} {\bf 11} (2015) 106,
  [\href{http://arxiv.org/abs/1504.07997}{{\tt arXiv:1504.07997}}].

\bibitem{Dolan:2003hv}
F.~A. Dolan and H.~Osborn, {\it {Conformal partial waves and the operator
  product expansion}},  {\em Nucl. Phys.} {\bf B678} (2004) 491--507,
  [\href{http://arxiv.org/abs/hep-th/0309180}{{\tt hep-th/0309180}}].

\bibitem{Brezin:1973aa}
A.~Brezin, J.~C. Le~Guillou, J.~Zinn-Justin, and B.~J. Nickel, {\it Higher
  order contributions to critical exponents},  {\em J. Phys.} {\bf A44} (1973),
  no.~3 227.

\bibitem{Braun:2013tva}
V.~M. Braun and A.~N. Manashov, {\it {Evolution equations beyond one loop from
  conformal symmetry}},  {\em Eur. Phys. J.} {\bf C73} (2013) 2544,
  [\href{http://arxiv.org/abs/1306.5644}{{\tt arXiv:1306.5644}}].

\end{thebibliography}\endgroup

\end{document}